\documentclass[12pt,english]{article}
\usepackage[OT1]{fontenc}
\usepackage[latin9]{inputenc}
\usepackage{geometry}
\usepackage{babel}
\usepackage{array}
\usepackage{textcomp}
\usepackage{mathtools}
\usepackage{bm}
\usepackage{multirow}
\usepackage{amsmath}
\usepackage{amssymb}
\usepackage{graphicx}
\usepackage{setspace}
\usepackage[numbers]{natbib}
\usepackage[unicode=true,pdfusetitle,
 bookmarks=true,bookmarksnumbered=false,bookmarksopen=false,
 breaklinks=false,pdfborder={0 0 1},backref=false,colorlinks=false]
 {hyperref}

\makeatletter

\providecommand{\tabularnewline}{\\}

\@ifundefined{date}{}{\date{}}
\usepackage[toctitles]{titlesec}
\usepackage{bm}

\usepackage[table]{xcolor}
\usepackage{makecell}
\renewenvironment{cases}{%
 \begin{dcases}%
}{%
 \end{dcases}
}

\def\XXint#1#2#3{{\setbox0=\hbox{$#1{#2#3}{\int}$ }
\vcenter{\hbox{$#2#3$ }}\kern-.6\wd0}}

\makeatother

\begin{document}

\title{Physical modelling of the slow voltage relaxation phenomenon in lithium-ion
batteries }

\author{Toby L. Kirk$^{1,2}$, Colin P. Please$^{1,2}$, S. Jon Chapman$^{1,2}$}

\maketitle
\vspace{-0.5cm}

\noindent\begin{minipage}[t]{1\columnwidth}%
\begin{center}
{\small{}\footnotemark[1]Mathematical Institute, University of Oxford,
}\\
{\small{}Andrew Wiles Building, Woodstock Road, Oxford, OX2 6GG, UK}{\small\par}
\par\end{center}
\begin{center}
{\small{}\footnotemark[2]The Faraday Institution, Quad One, Becquerel
Avenue,}\\
{\small{}Harwell Campus, Didcot, OX11 0RA, UK}{\small\par}
\par\end{center}%
\end{minipage}

\bigskip{}

\begin{abstract}
In the lithium-ion battery literature, discharges followed by a relaxation
to equilibrium are frequently used to validate models and their parametrizations.
Good agreement with experiment during discharge is easily attained
with a pseudo-two-dimensional model such as the Doyle-Fuller-Newman
(DFN) model. The relaxation portion, however, is typically not well-reproduced,
with the relaxation in experiments occurring much more slowly than
in models. In this study, using a model that includes a size distribution
of the active material particles, we give a physical explanation for
the slow relaxation phenomenon. This model, the Many-Particle-DFN
(MP-DFN), is compared against discharge and relaxation data from the
literature, and optimal fits of the size distribution parameters (mean
and variance), as well as solid-state diffusivities, are found using
numerical optimization. The voltage after relaxation is captured by
careful choice of the current cut-off time, allowing a single set
of physical parameters to be used for all C-rates, in contrast to
previous studies. We find that the MP-DFN can accurately reproduce
the slow relaxation, across a range of C-rates, whereas the DFN cannot.
Size distributions allow for greater internal heterogeneities, giving
a natural origin of slower relaxation timescales that may be relevant
in other, as yet explained, battery behavior.
\end{abstract}

\section{Introduction}

Lithium-ion batteries are rechargeable energy storage devices used
across the consumer electronics industry due to their long lifespan,
high energy density, and a low self-discharge rate compared to other
batteries \citep{Blomgren2017}. In recent years, their demand has
grown due to their use in electric vehicles, and it is predicted to
increase from 45 GWh per year in 2015 to 390 GWh per year by 2030
\citep{Zubi2018}, motivating improvements in battery performance.
Although experimental research is necessary to achieve this, mathematical
modelling also plays a key role\textemdash for recent reviews of the
various scales and complexities that have been modelled, see \citep{Franco2013,Ramadesigan2012}.
The pioneering continuum model, still used today, was developed by
the group of Newman \citep{DoyleFullerNewman1993,Fuller1994,Thomas2002,NewmanBook}
where a macroscale (i.e., cell scale) model is coupled at each location
to a microscale (i.e., particle scale) one\textemdash a picture justified
by asymptotic homogenization \citep{Richardson2012}, and also referred
to as pseudo-two-dimensional (P2D).

A frequently occurring scenario, both in battery use and in research,
is that of relaxation to equilibruim after a period of dynamic (dis)charging.
During a (dis)charge, the internal state of the cell is transient
with heterogeneities in lithium concentrations, in the electrolyte
and the active materials of both the positive and negative electrodes.
When the (dis)charge is stopped and the circuit current is switched
off, the internal states then equilibriate and relax to a uniform
steady state. This relaxation could take several hours, even up to
24 hours \citep{Roscher2011}, depending on the size of internal heterogeneities
at current cut-off. There are two key components of this relaxation
for a physical model to capture which are easily observed in experiments:
\begin{enumerate}
\item The final equilibrium voltage after relaxation;
\item The manner or shape of the voltage relaxation profile.
\end{enumerate}
Research on voltage relaxation has been focused mainly on improving
the accuracy of open circuit voltage (OCV) measurements, typically
done using the galvanostatic intermittent titration technique (GITT),
where steps are taken through states of charge (SoC) incrementally,
waiting for the cell to sufficiently relax after each step. The measurement
of property (1) is then of importance, and strategies have been developed
to measure this voltage accurately in a shorter time \citep{Li2016,Pei2014,Petzl2013},
achieved by fitting equivalent circuit models to (2) with several
(up to 5) RC elements, with large time constants that are difficult
to interpret physically \citep{Li2016}.

Battery relaxation has also found uses in the parametrization of physical
models. These include recent comprehensive studies by Ecker et al.
\citep{Ecker2015a,Ecker2015b}, Schmalstieg et al. \citep{Schmalstieg2018a,Schmalstieg2018b},
and Chen et al. \citep{Chen2020}, which use state-of-the-art experimental
characterization techniques to parametrize (variants of) the Doyle-Fuller-Newman
(DFN) model \citep{DoyleFullerNewman1993,Fuller1994} for commercial
cells. The DFN is used across the lithium-ion battery literature and
considered a benchmark. These studies use current pulses at a range
of SoCs \citep{Ecker2015a,Ecker2015b}\citep{Schmalstieg2018a,Schmalstieg2018b},
but also full discharges \citep{Ecker2015a,Ecker2015b} \citep{Schmalstieg2018a,Schmalstieg2018b}
\citep{Chen2020}, each followed by relaxations, in order to validate
their parametrizations and demonstrate the accuracy of their physical
models. However, the slow relaxations of the experiments, clearly
visible in \citep{Ecker2015b} (Figs. 5, 6, 8), \citep{Schmalstieg2018b}
(Figs. 6, 9), and \citep{Chen2020} (Fig. 17), are not captured by
the models, and are given minimal (if any) discussion. At low SoCs,
the final rest voltage is also highly inaccurate, as seen in \citep{Ecker2015b}
(Fig. 6), \citep{Schmalstieg2018b} (Figs. 6, 9). Chen et al. \citep{Chen2020}
use the final rest voltage after a discharge to inform their parameter
estimates, resulting in different parameter values needed for each
experiment. This slow relaxation is not confined to lithium ion batteries
but is also seen in other chemistries, e.g., lead acid where model
parametrizations have faced similar difficulties \citep{Sulzer2019b}.

In this paper, we will give a physical explanation for the slow relaxation
phenomenon using an extension of the DFN to include a distribution
of active particle sizes. Models with multiple (or distributions of)
particle sizes have been considered by several authors, e.g. \citep{Farkhondeh2011,Meyers2000,Song2013,Kirk2020},
but not in the context of relaxation. The model we use, denoted the
Many-Particle-DFN (MP-DFN), is compared against the discharge and
relaxation experimental data of \citep{Chen2020}, taken from a set
of commercial cells (LG M50). The MP-DFN is initially parametrized
by modifying the DFN parameter set in \citep{Chen2020}; the process
is described in sufficient detail to facilitate its use for other
cells and chemistries. Simulations are performed using the open source
software package Python Battery Mathematical Modelling (PyBaMM) \citep{Sulzer2020}
and optimal estimates of the microstructural parameters related to
the size distributions (mean and variance), and solid-state diffusivities,
are found by fitting the voltage profiles (across discharge and relaxation)
using a numerical optimization package (DFO-LS \citep{Cartis2019}).

The structure of the paper is as follows. The mathematical models
(MP-DFN and DFN), geometry and notation, are defined in section \ref{sec:Modelling};
the parameter set from the literature, and how it was adapted, is
in section \ref{sec:Parameter-values}. The methodology is described
in section \ref{sec:Methodology}, with the model comparisons (and
parameter fitting) presented in section \ref{sec:Results}, followed
by conclusions and future work.

\section{Modelling}

\label{sec:Modelling}

In this section, we describe the mathematical models considered in
this paper, the MP-DFN and DFN. We first describe the geometry and
notation used, then state the MP-DFN model as it is the most general,
followed by the DFN which is a special case of the MP-DFN. 

\subsection{Geometry and notation}

\global\long\def\k{\mathrm{k}}
\global\long\def\n{\mathrm{n}}
\global\long\def\sep{\mathrm{sep}}
\global\long\def\p{\mathrm{p}}
\global\long\def\e{\mathrm{e}}
\global\long\def\s{\mathrm{s}}

\begin{figure}
\begin{raggedright}
\includegraphics[width=0.85\textwidth]{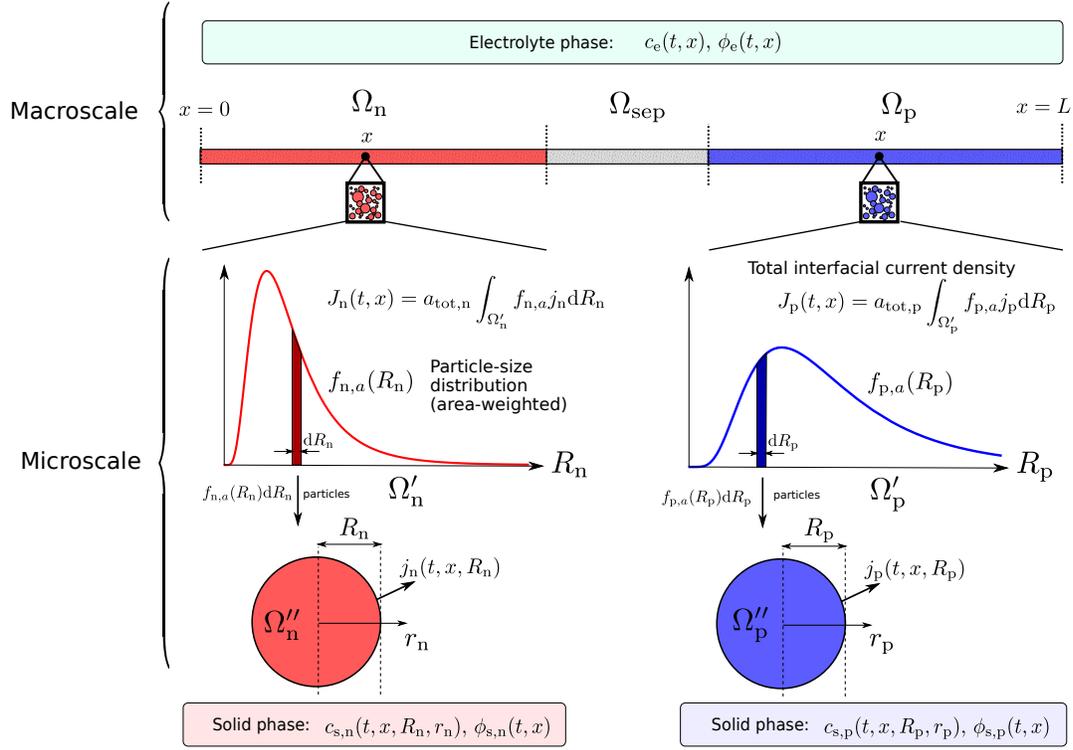}
\par\end{raggedright}
\caption{\label{fig:Schematic}Schematic depicting the multi-scale geometry
used for the Many-Particle-DFN model. The macroscale ($x$) and microscale
dimensions ($R_{\protect\k},r_{\protect\k}$ for $\protect\k=\protect\n,\protect\p$),
as well as the fundamental variables in the solid and electrolyte
phases, are shown.}

\end{figure}

First, we summarise the geometry and notation used within both models.
A schematic of the multiscale nature of the geometry is shown in Fig.
\ref{fig:Schematic}.The macroscale geometry of the cell is one dimensional,
with variation only in the through-cell direction, measured by the
coordinate $x$. Thicknesses of the negative electrode, separator
and positive electrode are $L_{\n}$, $L_{\sep}$, and $L_{\p}$,
respectively. The total thickness of the cell, from the negative current
collector (at $x=0$) to the positive current collector (at $x=L$)
is thus $L=L_{\n}+L_{\sep}+L_{\p}$. The macroscale is divided into
three regions,
\begin{align}
\Omega_{\n} & =\{0\leq x<L_{\n}\}, & \Omega_{\sep} & =\{L_{\n}\leq x<L-L_{\p}\}, & \Omega_{\p} & =\{L-L_{\p}\leq x<L\},\\
 & \text{(negative electrode)} &  & \text{(separator)} &  & \text{(positive electrode)}\nonumber 
\end{align}
At each macroscale location of both electrodes, $x\in\Omega_{\k}$,
$\k=\n,\p,$ there is a microscale domain $\Omega_{\k}^{\prime}=\{R_{\k,\mathrm{min}}\leq R_{\k}\leq R_{\k,\mathrm{max}}\}$
comprising a collection of spherical particles of solid active material.
The range of particle radii $R_{\k}$ that are present is modelled
as a continuum, taking all positive values between $R_{\k,\mathrm{min}}$
and $R_{\k,\mathrm{max}}$ (which could be $0$ and $\infty$), with
the fraction of all particles of a given radius $R_{\k}$ given by
the (normalised) particle-size distribution $f_{\k,n}(R_{\k})$. However,
it is more convenient to deal with the fraction of \emph{surface area}
contributed by particles of radius $R_{\k}$, which we denote $f_{\k,a}(R_{\k})$
and refer to as the area-weighted particle-size distribution (aPSD).
Particle size can then be interpreted as a microscale dimension, with
``coordinate'' $R_{\k}$. In this paper we consider the particle-size
distribution to be independent of macroscale location $x$, but one
could consider a non-uniform spatial distribution by letting $f_{\k,a}(R_{\k})$
(and also $\Omega_{\k}^{'}$, $R_{\k,\mathrm{min}}^{*}$, $R_{\k,\mathrm{max}}$,
etc.) depend on $x$.

All active particles of a given size (and at a given location) behave
identically, and have a further internal domain $\Omega_{\k}^{\prime\prime}(R_{\k})=\{0\leq r_{\k}\leq R_{\k}\}$
described (due to spherical symmetry) by the radial coordinate $r_{\k}$.
Hence, each electrode consists of a hierarchy of three domains or
dimensions\footnote{This means the model could be referred to as ``pseudo-three-dimensional''
or P3D, with particle size interpreted as another pseudo-dimension.},
\begin{equation}
\widehat{\Omega}_{\k}=\bigcup_{x\in\Omega_{\k}}\left[\bigcup_{R_{\k}\in\Omega_{\k}^{\prime}}\Omega_{\k}^{\prime\prime}(R_{\k})\right].
\end{equation}
We use the subscript $\k\in\{\n,\sep,\p\}$ to indicate in which subdomain
that variable is defined. Then for that subdomain, the phase, either
solid or electrolyte, is denoted by the additional subscript $\s$
or $\e$, respectively. The variables in the model and their subdomains
are
\begin{align}
\text{Electrolyte phase : } & \phi_{\e,\k},c_{\e,\k},i_{\e,\k},N_{\e,\k} & x\in\Omega_{\k}, & \,\,\k=\n,\sep,\p\\
\text{Solid phase : } & \phi_{\s,\k},i_{\s,\k} & x\in\Omega_{\k}, & \,\,\k=\n,\p\\
 & c_{\s,\k},N_{\s,\k} & x\in\Omega_{\k},\,R_{\k}\in\Omega_{\k}^{\prime},\,r_{\k}\in\Omega_{\k}^{\prime\prime}(R_{\k}), & \,\,\k=\n,\p
\end{align}
where potentials are denoted by $\phi$, current densities by $i$,
molar lithium concentrations by $c$ (with $c_{\e,\k}$ being lithium-ion
concentrations), and molar fluxes by $N$. We note that all quantities
depend on the macroscale coordinate $x$ and time $t$, but $c_{\s,\k}$
(and $N_{\s,\k}$) depend additionally on the microscale coordinates:
particle radius $R_{\k}\in[R_{\k,\mathrm{min}},R_{\k,\mathrm{max}}]$
and the radial coordinate $r_{\k}\in[0,R_{\k}]$. The parameters are
described in Table \ref{tab:Dimensional-parameters}, along with their
values from the literature \citep{Chen2020}.

\subsection{Many-Particle-Doyle-Fuller-Newman model (MP-DFN)}

\subsubsection{Dimensional governing equations}

\subsubsection*{Charge conservation}

The conservation of charge in the electrolyte and electrode phases
is given by

\begin{align}
\frac{\partial i_{\e,\k}}{\partial x} & =-\frac{\partial i_{\s,\k}}{\partial x}=\begin{cases}
J_{\k}, & \k=\n,\p,\\
0, & \k=\sep,
\end{cases} & \k\in\{\n,\sep,\p\}.\label{eq:dimensional_charge_conservation}
\end{align}
The interfacial current density $J_{\k}$ represents the total charge
transfer, due to electrochemical reactions, between the active material
and electrolyte at a given $x$ location. The current densities in
the electrolyte and electrode material are given by MacInnes' equation
and Ohm's law, respectively,
\begin{equation}
i_{\e,\k}=\epsilon_{\k}^{b_{\k}}\kappa_{\e}(c_{\e,\k})\left[-\frac{\partial\phi_{\e,\k}}{\partial x}+2(1-t^{+})\frac{R_{g}T}{F}\frac{\partial}{\partial x}\log c_{\e,\k}\right],\qquad\k\in\{\n,\sep,\p\},
\end{equation}
\begin{equation}
i_{\s,\k}=-\sigma_{\k}\frac{\partial\phi_{\s,\k}}{\partial x},\qquad\k\in\{\n,\p\}.
\end{equation}
There is continuity of $i_{\e,\k},\phi_{\e,\k}$ and $i_{\s,\k}=0$
at the internal electrode/separator boundaries, $x=L_{\n},L-L_{\p}$.
At the current collectors, charge only enters/exits the cell via the
solid phase, with current density $i_{\s,\n}=i_{\mathrm{app}}$ (time-dependent,
in general) imposed at $x=0$, but also $i_{\s,\p}=i_{\mathrm{app}}$
at $x=L$ by conservation of charge:
\begin{align}
i_{\e,\k} & =0,\qquad\text{at }x=0,L\\
i_{\s,\k} & =i_{\mathrm{app}}(t),\qquad\text{at }x=0,L
\end{align}
Then the solid-phase potentials at the current collectors are
\begin{align}
\phi_{\s,\n} & =0\quad\text{at }x=0, & \phi_{\s,\p} & =V\quad\text{at }x=L.
\end{align}
Adding the $i_{\e,\k}$ and $i_{\s,\k}$ equations in (\ref{eq:dimensional_charge_conservation}),
integrating over each region and imposing continuity gives that $i_{\s,\k}+i_{\e,\k}=i_{\mathrm{app}}$
in each electrode which can be used to eliminate $i_{\s,\k}$, and
in the separator,
\begin{equation}
i_{\e,\sep}\equiv i_{\mathrm{app}}.
\end{equation}

\subsubsection*{Molar conservation of lithium}

In the electrolyte:

\begin{align}
\epsilon_{\k}\frac{\partial c_{\e,\k}}{\partial t} & =-\frac{\partial N_{\e,\k}}{\partial x}+\frac{1}{F}\frac{\partial i_{\e,\k}}{\partial x},\quad\k\in\{\n,\sep,\p\},\label{eq:dimensional_lithium_cons_electrolyte}\\
N_{\e,\k} & =-\epsilon_{\k}^{b_{\k}}D_{\e}(c_{\e,\k})\frac{\partial c_{\e,\k}}{\partial x}+\frac{t^{+}}{F}i_{\e,\k},\quad\k\in\{\n,\sep,\p\},
\end{align}
with continuity of $c_{\e,\k},N_{\e,\k}$ at internal boundaries $x=L_{\n},L-L_{\p}$,
and no-flux $N_{\e,\k}=0$ at the current collectors $x=0,L$.

In the active solid electrode particles, lithium transport is modelled
by Fickian diffusion,

\begin{align}
\frac{\partial c_{\s,\k}}{\partial t} & =-\frac{1}{r_{\k}^{2}}\frac{\partial}{\partial r_{\k}}(r_{\k}^{2}N_{\s,\k}),\quad\k\in\{\n,\p\},\\
N_{\s,\k} & =-D_{\s,\k}\frac{\partial c_{\s,\k}}{\partial r_{\k}},\quad\k\in\{\n,\p\},
\end{align}
with regularity at the particle centres, and a flux condition at the
surface,
\begin{align}
N_{\s,\k} & =0,\quad\text{at }r_{\k}=0, & N_{\s,\k} & =\frac{j_{\k}}{F},\quad\text{at }r_{\k}=R_{\k},
\end{align}
where $j_{\k}$ is the interfacial current density representing the
charge transfer into the electrolyte, which may be different for particles
of different sizes. The total interfacial current density originating
from all particles at that macroscale location $x$ is then
\begin{equation}
J_{\k}=\int_{\Omega_{\k}^{'}}a_{\k}(R_{\k})j_{\k}\,\mathrm{d}R_{\k}=a_{\mathrm{tot},\k}\int_{\Omega_{\k}^{'}}f_{\k,a}(R_{\k})j_{\k}\,\mathrm{d}R_{\k}\label{eq:dimensional_J_k}
\end{equation}
which appears as a charge source/sink in (\ref{eq:dimensional_charge_conservation}),
and a lithium-ion source/sink in (\ref{eq:dimensional_lithium_cons_electrolyte}).
The total active surface area per unit volume, $a_{\mathrm{tot},\k}$,
is determined from the volume fraction of active material, $\epsilon_{s,\k}$,
and the distribution $f_{\k,a}$ \citep{Kirk2020}
\begin{equation}
a_{\mathrm{tot},\k}=\frac{3\epsilon_{s,\k}}{\int_{\Omega_{\k}^{'}}R_{\k}f_{\k,a}(R_{\k})\,\mathrm{d}R_{\k}}=\frac{3\epsilon_{s,\k}}{\bar{R}_{\k,a}},\label{eq:a_k_star}
\end{equation}
where the factor of 3 is due to the assumption here of spherical particles.
The radius $\bar{R}_{\k,a}$ is the area-weighted mean radius, or
the mean of $f_{\k,a}(R_{\k})$\textemdash see section \ref{subsec:Microscale-parameters}
for further discussion of aPSD quantities and their relevance.

\subsubsection*{Electrochemical reactions}

The interfacial current density originating from the lithium (de)intercalation
reaction on the surface ($r_{\k}=R_{\k})$ of all active particles
in each electrode is modelled by symmetric Butler\textendash Volmer
kinetics (transfer coefficients equal to 1/2): 
\begin{align}
j_{\k} & =j_{\k,0}\sinh\left(\frac{F\eta_{\k}}{2R_{g}T}\right),\qquad\k\in\{\n,\p\},\label{eq:Butler-Volmer_dim}\\
\text{Exchange current density : }j_{0,\k} & =m_{\k}(c_{\s,\k})^{1/2}(c_{\k,\mathrm{max}}-c_{\s,\k})^{1/2}(c_{\e,\k})^{1/2},\qquad\k\in\{\n,\p\},\label{eq:Exchange_current_density_dim}\\
\text{Reaction overpotential : }\eta_{\k} & =\phi_{\s,\k}-\phi_{\e,\k}-U_{\k}(c_{\s,\k}),\qquad\k\in\{\n,\p\},\label{eq:Overpotential_dim}
\end{align}
In the above it is implicit that the concentration $c_{\s,\k}$ is
evaluated on the particle surface $r_{\k}=R_{\k}$.

\subsubsection*{Initial conditions}

Initially, at $t=0$, we take the cell to be at rest with all variables
constant and uniform in space. The initial conditions are
\begin{align}
c_{\s,\k} & =c_{\k,0},\qquad\k\in\{\n,\p\},\\
c_{\e,\k} & =c_{\e,0},\qquad\k\in\{\n,\sep,\p\},
\end{align}
which forces the initial potentials to be
\begin{align}
U_{\k} & =U_{\k}(c_{\k,0})=U_{\k,0},\qquad\k\in\{\n,\p\},\\
\phi_{\s,\k} & =\begin{cases}
0, & \k=\n,\\
U_{\p,0}-U_{\n,0} & \k=\p,
\end{cases}\\
\phi_{\e,\k} & =-U_{\n,0},\qquad\k\in\{\n,\sep,\p\},
\end{align}
and all other variables are initially equal to zero.

\subsection{Doyle-Fuller-Newman model}

\label{subsec:DFN-model}

Here we state the standard Doyle-Fuller-Newman (DFN) model, the most
commonly used physical porous electrode model of a lithium-ion cell,
and the model employed in Chen et al. \citep{Chen2020}. If we set
all the particles in electrode $\k$ to be the same size, $R_{\k}=R_{\k,DFN}$,
then the aPSD is a Dirac delta function, $f_{\k,a}(R_{\k})=\delta(R_{\k}-R_{\k,DFN})$,
and the MP-DFN model reduces to the DFN model, with equation (\ref{eq:dimensional_J_k})
reducing to $J_{\k}=a_{\k}j_{\k}|_{R_{\k}=R_{\k,DFN}}$. Then, only
the dynamics of particles of size $R_{\k}=R_{\k,DFN}$ need to be
modelled, and hence all of the remaining equations are identical to
that of the MP-DFN, but with all variables (where applicable) evaluated
at $R_{\k}=R_{\k,DFN}$.

\section{\label{sec:Parameter-values}Parameter values from the literature}

In order to later compare to the experimental results of Chen et al.
\citep{Chen2020} we use a parameter set based on their extensive
parametrization of a cylindrical 21700 commercial cell (LGM50), tailored
to their P2D model. The model in Chen et al. \citep{Chen2020} is
identical to the DFN, given in section \ref{sec:Modelling} up to
differences in notation. However, as the DFN assumes only a single
particle size for each electrode, several modifications or additions
to the parameter set are needed before its use with the MP-DFN, which
we now detail.

\begin{table*}
\begin{centering}
\begin{tabular}{clcccc}
\hline 
{\footnotesize{}Dimensional} & \multirow{2}{*}{{\footnotesize{}Description {[}unit{]}}} & \multicolumn{3}{c}{{\footnotesize{}Value, region $\k$}} & \multirow{2}{*}{{\footnotesize{}Ref.}}\tabularnewline
{\footnotesize{}parameter} &  & {\footnotesize{}$\n$} & {\footnotesize{}$\sep$} & {\footnotesize{}$\p$} & \tabularnewline
\hline 
$R_{g}$ & {\footnotesize{}Universal gas constant $[\mathrm{J\,mol}^{-1}\mathrm{K}^{-1}]$} & \multicolumn{3}{c}{{\footnotesize{}8.3145}} & \multirow{15}{*}{{\footnotesize{}\citep{Chen2020}}}\tabularnewline
$F$ & {\footnotesize{}Faraday's constant $[\mathrm{Cmol}^{-1}]$} & \multicolumn{3}{c}{{\footnotesize{}96485}} & \tabularnewline
$T$ & {\footnotesize{}Temperature $[\mathrm{K}]$} & \multicolumn{3}{c}{{\footnotesize{}298.15}} & \tabularnewline
$C$ & {\footnotesize{}\makecell[l]{Reference current density\\to discharge in 1hr [Am$^{-2}$]}} & \multicolumn{3}{c}{{\footnotesize{}5}} & \tabularnewline
$c_{\e,0}$ & {\footnotesize{}\makecell[l]{Initial Li concentration\\in electrolyte  [mol m$^{-3}$]}} & \multicolumn{3}{c}{{\footnotesize{}1000}} & \tabularnewline
$D_{\e,\mathrm{typ}}$ & {\footnotesize{}\makecell[l]{Typical diffusivity of lithium\\ions in electrolyte [m$^2$s$^{-1}$]}} & \multicolumn{3}{c}{{\footnotesize{}$1.77\times10^{-10}$}} & \tabularnewline
{\footnotesize{}$\kappa_{\e,\mathrm{typ}}$} & {\footnotesize{}\makecell[l]{Typical conductivity\\of electrolyte [S m$^{-1}$]}} & \multicolumn{3}{c}{{\footnotesize{}0.949}} & \tabularnewline
{\footnotesize{}$t^{+}$} & {\footnotesize{}Cation transference number} & \multicolumn{3}{c}{{\footnotesize{}0.2594 }} & \tabularnewline
$\epsilon_{\k}$ & {\footnotesize{}Electrolyte volume fraction} & {\footnotesize{}0.25} & {\footnotesize{}0.47} & {\footnotesize{}0.335 } & \tabularnewline
$b_{\k}$ & {\footnotesize{}Bruggeman coefficient} & {\footnotesize{}1.5} & {\footnotesize{}1.5} & {\footnotesize{}1.5} & \tabularnewline
$L_{\k}$ & {\footnotesize{}Electrode thickness $[\mathrm{m}]$} & {\footnotesize{}$85.2\times10^{-6}$} & {\footnotesize{}$12\times10^{-6}$} & {\footnotesize{}$75.6\times10^{-6}$} & \tabularnewline
$\epsilon_{\s,\k}$ & {\footnotesize{}Active material volume fraction} & {\footnotesize{}0.75} & {\footnotesize{}-} & {\footnotesize{}0.665 } & \tabularnewline
{\footnotesize{}$\sigma_{\k}$} & {\footnotesize{}\makecell[l]{Conductivity in \\in electrode material [S m$^{-1}$]}} & {\footnotesize{}215 } & {\footnotesize{}-} & {\footnotesize{}0.18} & \tabularnewline
$c_{\k,\mathrm{max}}$ & {\footnotesize{}\makecell[l]{Max. Li concentration in\\active material [mol m$^{-3}$]}} & {\footnotesize{}33133 } & {\footnotesize{}-} & {\footnotesize{}63104 } & \tabularnewline
$c_{\k,0}$ & {\footnotesize{}\makecell[l]{Initial Li concentration in\\active material [mol m$^{-3}$]}} & {\footnotesize{}29866} & {\footnotesize{}-} & {\footnotesize{}17038} & \tabularnewline
\hline 
$U_{\k}(c_{s,\k})$ & {\footnotesize{}\makecell[l]{Open circuit potential\\relative to Li/Li$^+$ [V]}} & {\footnotesize{}Eq. (\ref{eq:OCP_fit_n_dim})} & {\footnotesize{}-} & {\footnotesize{}Eq. (\ref{eq:OCP_fit_p_dim})} & \multirow{7}{*}{{\footnotesize{}\makecell[l]{Modified\\from \cite{Chen2020}}}}\tabularnewline
$m_{\k}$ & {\footnotesize{}Reaction rate $[\mathrm{A\,m}^{-2}(\mathrm{m}^{3}/\mathrm{mol})^{1.5}]$} & {\footnotesize{}$8.053\times10^{-7}$} & {\footnotesize{}-} & {\footnotesize{}$4.443\times10^{-6}$} & \tabularnewline
$D_{\s,\k}$ & {\footnotesize{}Diffusivity of Li in electrode $[\mathrm{m}^{2}\mathrm{s}^{-1}]$} & {\footnotesize{}$5.10\times10^{-14}$} & {\footnotesize{}-} & {\footnotesize{}$6.75\times10^{-15}$} & \tabularnewline
$f_{\k,a}(R_{\k})$ & {\footnotesize{}\makecell[l]{Area-weighted particle-size\\distribution (aPSD) [m$^{-1}$]}} & {\footnotesize{}Eq. }(\ref{eq:lognormal}) & {\footnotesize{}-} & {\footnotesize{}Eq. }(\ref{eq:lognormal}) & \tabularnewline
$\bar{R}_{\mathrm{\k},a}$ & {\footnotesize{}\makecell[l]{Area-weighted mean\\particle radius [m]}} & {\footnotesize{}$7.28\times10^{-6}$} & {\footnotesize{}-} & {\footnotesize{}$6.78\times10^{-6}$} & \tabularnewline
$\sigma_{\k,a}$ & {\footnotesize{}\makecell[l]{Area-weighted particle-\\size standard deviation [m]}} & {\footnotesize{}$2.08\times10^{-6}$} & {\footnotesize{}-} & {\footnotesize{}$2.59\times10^{-6}$} & \tabularnewline
$a_{\mathrm{tot},\mathrm{\k}}$ & {\footnotesize{}\makecell[l]{Total active surface area\\per volume from (\ref{eq:a_k_star}) [m$^{-1}$]}} & {\footnotesize{}$3.09\times10^{5}$} & {\footnotesize{}-} & {\footnotesize{}$2.94\times10^{5}$} & \tabularnewline
\hline 
\end{tabular}
\par\end{centering}
\caption{Dimensional parameters for use with the MP-DFN and DFN models, from
Chen et al. \citep{Chen2020}, with modified or new parameters indicated\textemdash see
section \ref{sec:Parameter-values}. Initial guesses used for the
fitting of section \ref{sec:Results}.}
\label{tab:Dimensional-parameters}
\end{table*}

\subsection{Microscale parameters}

\label{subsec:Microscale-parameters}

\begin{figure}
\includegraphics[width=1\textwidth]{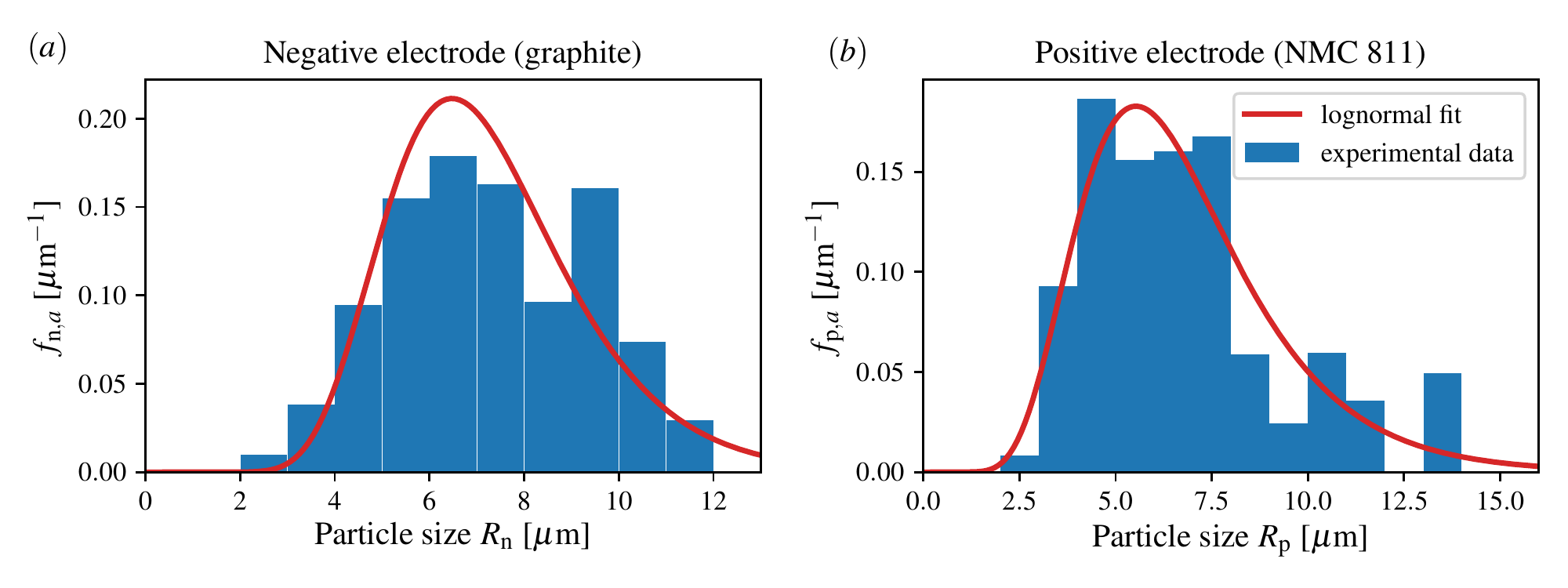}

\caption{\label{fig:aPSD_fits}Experimental measurements and lognormal fits
of area-weighted particle-size distributions $f_{\protect\k,a}(R_{\protect\k})$,
using data from Chen et al. \citep{Chen2020}. (a) Negative electrode
$\protect\k=\protect\n$ (graphite only); (b) Positive electrode $\protect\k=\protect\p$
(NMC 811).}

\end{figure}

The MP-DFN requires not only a mean or representative particle radius
for each electrode, but a particle-size distribution. Particle-size
distributions were measured in Chen et al. \citep{Chen2020} using
scanning electron microscopy (SEM) images of electrode cross-sections
and grouping the particles of similar sizes to produce histograms
with bin widths of 1 $\mu$m (NMC 811 and graphite) or 0.5 $\mu$m
(SiO$_{x}$). This amounts to a (discrete) estimate of the number
density $n_{\k}(R_{\k})$, i.e., the number of particles of size $R_{\k}$
per unit volume in electrode $\k$. It is more convenient, but also
more physically relevant due to the interfacial nature of the electrochemical
reactions (see \citep{Kirk2020}), to work in terms of the area density
$a_{\k}(R_{\k})=4\pi R_{\k}^{2}n_{\k}(R_{\k})$, or its normalised
version, the area-weighted particle-size distribution (aPSD), 
\begin{align}
f_{\k,a}(R_{\k}) & =\frac{a_{\k}(R_{\k})}{\int_{\Omega_{\k}^{'}}a_{\k}(R_{\k})\,\mathrm{d}R_{\k}}=\frac{a_{\k}(R_{\k})}{a_{\mathrm{tot},\k}}\\
 & =\frac{R_{\k}^{2}n_{\k}(R_{\k})}{\int_{\Omega_{\k}^{'}}R_{\k}^{2}n_{\k}(R_{\k})\,\mathrm{d}R_{\k}}.\label{eq:n_to_f_a}
\end{align}
For the MP-DFN, it is sufficient to specify the function $a_{\k}(R_{\k})$,
which we achieve by specifying its integral, $a_{\mathrm{tot},\k}$,
and its shape, $f_{\k,a}(R_{\k})$ (which integrates to one). Here
we convert the measurements of $n_{\k}(R_{\k})$ in Chen et al. \citep{Chen2020}
to a discrete area-weighted (and normalised) distribution using (\ref{eq:n_to_f_a}),
and then fit a continuous density function $f_{\k,a}(R_{\k})$. The
distribution data is shown in Fig. \ref{fig:aPSD_fits} for the positive
electrode (NMC) and negative electrode (graphite, with the small contributions
of SiO$_{x}$ neglected). The distributions are unimodal with positive
skew, motivating their representation by a lognormal\textemdash commonly
used for electrode PSDs \citep{Meyers2000,Song2013,Kirk2020}\textemdash given
by
\begin{align}
f_{\k,a}(R) & =\frac{1}{R\sqrt{2\pi\sigma_{\k,LN}^{2}}}\exp\left[-\frac{(\log R-\mu_{\k,LN})^{2}}{2\sigma_{\k,LN}^{2}}\right],\qquad\qquad\k\in\{\n,\p\},\label{eq:lognormal}
\end{align}
with shape parameters $\mu_{\k,LN}\in(-\infty,\infty),\,\sigma_{\k,LN}>0,$
related to the mean and variance via
\begin{align}
\bar{R}_{\k,a} & =\exp(\mu_{\k,LN}+\sigma_{\k,LN}^{2}/2),\\
\sigma_{\k,a}^{2} & =\exp(\sigma_{\k,LN}^{2}-1)\exp(2\mu_{\k,LN}+\sigma_{\k,LN}^{2}),
\end{align}
Least-square fits of lognormals to the data, using the SciPy optimize
package, are also shown in Fig. \ref{fig:aPSD_fits}, and the distribution
parameters (mean $\bar{R}_{\k,a}$ and standard deviation $\sigma_{\k,a}$)
are given in Table \ref{tab:Dimensional-parameters}. By our choice
of distribution family, the result is one additional parameter per
electrode, the area-weighted standard deviation $\sigma_{\k,a}$,
not present in the DFN.

Given the lognormal fits for $f_{\k,a}(R_{\k})$ and the active material
volume fraction $\epsilon_{\s,\k}$ (taken from \citep{Chen2020}
unaltered), the total surface area per unit volume is then calculated
via (\ref{eq:a_k_star})\textemdash see \citep{Kirk2020}. Note that
(\ref{eq:a_k_star}) is similar to that typically used in the literature
for the DFN, where all particles are of a single radius, but with
the appropriate single radius given by the area-weighted mean $\bar{R}_{\k,a}$.
In Chen et al. \citep{Chen2020}, the number-based mean particle radii
were used, which are less than the fitted values of $\bar{R}_{\k,a}$
as calculated here. Therefore, given the active material volume, \citep{Chen2020}
overestimates the surface areas compared to the true areas calculated
using the PSD.

\subsection{Other modified parameters}

The determination of several parameters in the set provided in \citep{Chen2020}
relied on an estimate of the typical or single representative particle
size for each electrode. These parameters include the reaction rates
($m_{\k},\ \k=\n,\p$) and solid-state lithium diffusion coefficients
($D_{\s,\k},\ \k=\n,\p$), determined via electrochemical impedance
spectroscopy (EIS) and pulse galvanostatic intermittent titration
technique (pulse GITT), respectively. Chen et al. \citep{Chen2020}
used the number-based mean radius, $\bar{R}_{\mathrm{\k},n}$, but
here we choose the area-weighted mean radius, $\bar{R}_{\mathrm{\k},a}$.
As shown in Kirk et al. \citep{Kirk2020}, a sphere of radius $\bar{R}_{\mathrm{\k},a}$
exhibits precisely the same surface-area-to-volume ratio as the particle
population. This mean radius is unique in this respect, and is therefore
a much better choice to represent the PSD \citep{Kirk2020}. The aforementioned
reaction rates\footnote{In the PyBaMM implementation, the parameter that is rescaled in practice
is the exchange current density $j_{0,\k}$.} and diffusion coefficients based on $\bar{R}_{\mathrm{\k},n}$ can
be readily updated so that they are instead based on $\bar{R}_{\mathrm{\k},a}$
by a simple rescaling, and thus provide a consistent parameter set
for the MP-DFN. The reaction rates $m_{\k}$ were determined from
EIS data using Eqs (18) and (20) in \citep{Chen2020}, from which
we find $m_{\k}\propto(a_{\mathrm{tot},\k})^{-1}\propto\bar{R}_{\mathrm{\k},a}$.
The diffusion coefficients $D_{\s,\k}$ were determined using Eq.
(14) in Chen et al., following GITT pulses (averaged across experiments
at different states of charge, then later tuned for different C-rates),
from which we observe $D_{\s,\k}\propto(a_{\mathrm{tot},\k})^{-2}\propto(\bar{R}_{\mathrm{\k},a})^{2}$.
Hence, estimates based on the new radius can be found via the transformations
\begin{align}
m_{\k} & =\left(\frac{\bar{R}_{\mathrm{\k},a}}{\bar{R}_{\mathrm{\k},n}}\right)m_{\k}^{\mathrm{Chen}}, & D_{\s,\k} & =\left(\frac{\bar{R}_{\mathrm{\k},a}}{\bar{R}_{\mathrm{\k},n}}\right)^{2}D_{\s,\k}^{\mathrm{Chen}},
\end{align}
where $\bar{R}_{\mathrm{\k},a}/\bar{R}_{\mathrm{\k},n}=1.24$, $1.30$
for $\k=\n,\p$, and $m_{\k}^{\mathrm{Chen}}$, $D_{\s,\k}^{\mathrm{Chen}}$
are the Chen et al. \citep{Chen2020} estimates. The resulting new
estimates are given in Table \ref{tab:Dimensional-parameters}.

An important observation is that these updated parameter estimates
are not just applicable to the MP-DFN, but also the DFN if the particle
radius $\bar{R}_{\mathrm{\k},a}$ is used. This is because the relevant
timescales in the model\textemdash the diffusion timescales $\tau_{\k}$
and reaction timescales $\tau_{\mathrm{r},\k}$\textemdash are unchanged
from Chen et al, which is expected since these are the quantities
they directly measured experimentally. As a consequence, the dimensionless
DFN model equations are the same for both parameter sets, the original
set from \citep{Chen2020} and our modified one, resulting in the
same terminal voltage and current response.

\begin{figure}
\includegraphics[width=1\textwidth]{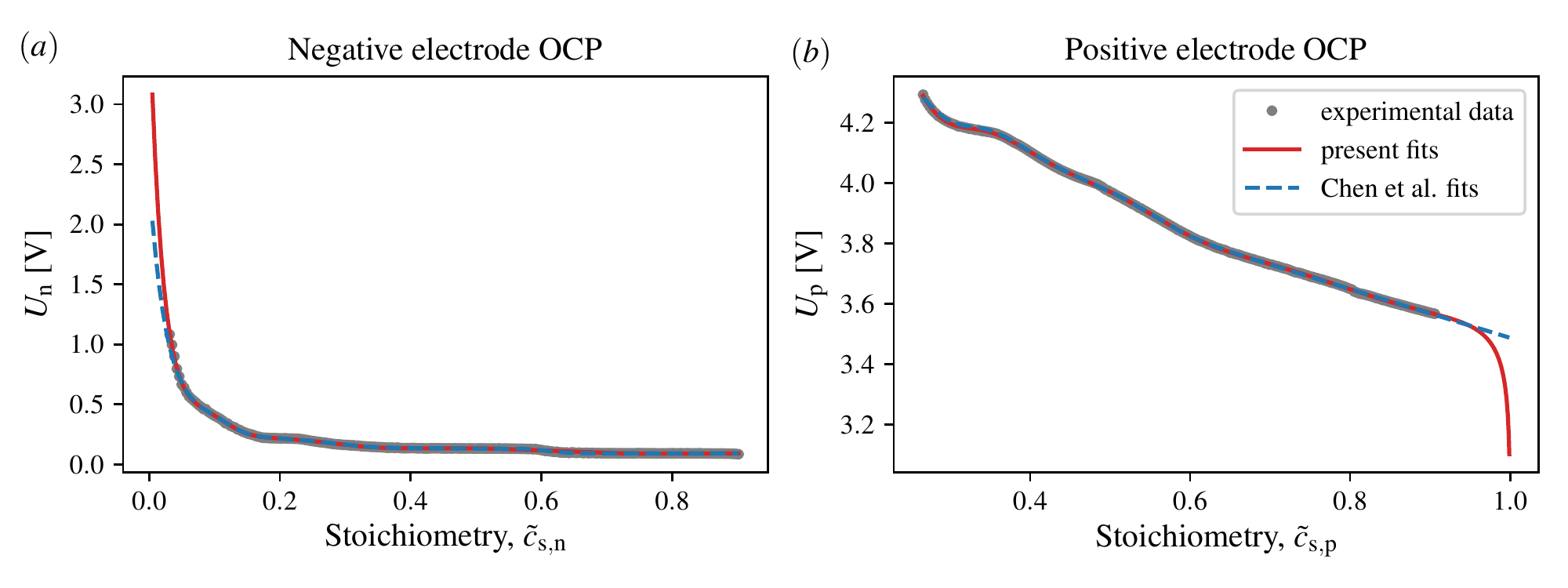}

\caption{\label{fig:OCP-fits}OCP analytical fits to OCP data from \citep{Chen2020}.
Shown are the fits from Chen et al. \citep{Chen2020} (Eqs. 8 and
9 therein), and our fits (Eqs. (\ref{eq:OCP_fit_n_dim}) and (\ref{eq:OCP_fit_p_dim}))
that include logarithmic singularities at $\tilde{c}_{\protect\s,\protect\k}=0,1$.
($a$) Negative electrode $\protect\k=\protect\n$ (graphite-SiO$_{x}$);
($b$) Positive electrode $\protect\k=\protect\p$ (NMC 811).}
\end{figure}

Lastly, it was necessary to slightly modify the fitted functional
forms of the OCPs, i.e. $U_{\k}(\tilde{c}_{\s,\k})$ where $\tilde{c}_{\s,\k}=c_{\s,\k}/c_{\k,\mathrm{max}}$
is the stoichiometry. Analytical fits were preferable to interpolation
of the OCP data of \citep{Chen2020} for two reasons: (i) to reduce
computation time given the high dimensionality and complexity of the
MP-DFN; (ii) to reduce erratic behaviour in smaller particles due
to experimental noise, as they are more sensitive to small deviations
in the OCP. Chen et al. \citep{Chen2020} provides smooth analytical
fits, but they were insufficient for our purposes since the expected
logarithmic singular behaviour of $U_{\k}(\tilde{c}_{\s,\k})$ at
electrode depletion and saturation ($\tilde{c}_{\s,\k}\to0,1$), preventing
the stoichiometries going below zero or above one, was not accounted
for in the functional forms. The lack of any singularity was an issue
particularly in the positive electrode because the experimental data
is only provided up to a stoichiometry of $\tilde{c}_{\s,\p}\approx0.91$,
with no visible indication of a singularity at $\tilde{c}_{\s,\p}=1$.
This range is sufficient for the DFN, since the stoichiometry in the
particle of mean radius remains less than 0.91 for a discharge to
the cut-off voltage of $2.5$ V. However, in the MP-DFN, smaller particles
lithiate (or delithiate) more quickly and reach stoichiometries much
closer to 1 (or 0), where they are prevented from exceeding 1 (or
0) by the steep gradients in the OCP. Therefore, we modified the analytical
OCPs to include theoretical logarithmic terms consistent with the
Butler\textendash Volmer equation (\ref{eq:Butler-Volmer_dim})-(\ref{eq:Overpotential_dim})
with transfer coefficients of $1/2$ (see e.g. \citep{PlettBookVol1}),
and then refitted the functions to the OCP data (available in PyBaMM).
The resulting fits, using the SciPy Optimize package, are given by

\begin{align}
U_{\n}(c)=\, & \frac{1}{2}\frac{R_{g}T}{F}\log\left(\frac{1-c}{c}\right)+3.5392\,\exp(-50.381c)-0.13472\nonumber \\
 & +98.941\tanh[4.1465(c-0.33873)]+102.43\tanh[3.8043(c-0.31895)]\nonumber \\
 & -0.19988\tanh[22.515(c-0.11667)]-200.87\tanh[3.9781(c-0.32969)],\label{eq:OCP_fit_n_dim}\\
U_{\p}(c)= & \,5\frac{R_{g}T}{F}\log\left(\frac{1-c}{c}\right)-27.648\,c+52.167\nonumber \\
 & -56.030\tanh[6.7733(c-0.53398)]+57.409\tanh[6.7071(c-0.53334)]\nonumber \\
 & +53.227\tanh[0.67406(c-1.6653)]+0.49701\tanh[14.355(c-0.30713)],\label{eq:OCP_fit_p_dim}
\end{align}
and are shown in Fig. \ref{fig:OCP-fits}. They accurately fit the
experimental data over the stoichiometry range provided, with RMSEs
of 4.19 mV and 1.48 mV, respectively, but they also exhibit the appropriate
singular behaviour near $\tilde{c}_{\s,\n}=0$ (Fig. \ref{fig:OCP-fits}($a$))
and $\tilde{c}_{\s,\p}=1$ (Fig. \ref{fig:OCP-fits}($b$)). We remark
that the singularity strength (numerical coefficient of the log) in
(\ref{eq:OCP_fit_p_dim}) was increased from $1/2$ to 5 in order
to improve the robustness of the parameter fitting procedure of section
(\ref{sec:Results}), allowing a wider range of parameters (i.e.,
those far from optimal) to be explored without violating the stoichiometry
limits during the simulations. The impact on the model behaviour at
the resulting optimal parameters, however, was minimal. 

\section{Methodology}

\label{sec:Methodology}

\subsection{Numerical simulations}

The PyBaMM \citep{Sulzer2020} package was used for simulations of
the DFN and MP-DFN models. PyBaMM is an open source software that
can quickly and robustly solve a variety of continuum battery models
in a modular or ``plug and play'' framework. The DFN model was already
available in PyBaMM, but the MP-DFN was newly implemented for the
present work. The implementation is will be made available in a future release of PyBaMM. For the results
presented here, we employed finite volume discretizations in each
spatial domain, with 30 volumes in each particle, 20 volumes (or ``size
bins'') in the particle-size dimensions, 20 volumes in each electrode
and 20 volumes in the separator. Finer meshes (up to 50, 50, 80 and
80 volumes in the respective domains) were also considered but the
difference from the results presented here were negligible. The discretization
results in a system of differential algebraic equations, and the time
integration was performed using the fast CasADi solver \citep{Andersson2019}
(employing automatic differentiation and the SUNDIALS IDA \citep{hindmarsh2005sundials}
package for systems of DAEs written in C) which is conveniently interfaced
directly from PyBaMM. In total, the MP-DFN system consisted of at
least 24061 ordinary differential equations and 100 algebraic equations,
and the absolute and relative tolerances of the solver were taken
to be their default values of $10^{-6}$. Each simulation of discharge
plus relaxation took on the order of 40-50 s on a laptop computer
with an Intel\textregistered{} Core i5-8350U CPU (1.70GHz \texttimes{}
8) and 16 GB RAM.

Regarding the particle-size dimensions $R_{\k}$, particular to the
present work, it was also numerically necessary to impose minimum
and maximum particle radii, $R_{\k,\mathrm{min}}$ and $R_{\k,\mathrm{max}}$,
for each electrode $\k=\n,\p$. We chose values based on the microscale
parameters in Table \ref{tab:Dimensional-parameters} with $R_{\k,\mathrm{min}}^{*}=0.1\bar{R}_{\k,a}$
and $R_{\k,\mathrm{max}}^{*}=6\bar{R}_{\k,a}$ found to be sufficiently
small and large (relative to the mean), respectively, giving $[R_{\n,\mathrm{min}},R_{\n,\mathrm{max}}]=[0.728\,\mu\mathrm{m},43.68\,\mu\mathrm{m}]$
and $[R_{\p,\mathrm{min}},R_{\p,\mathrm{max}}]=[0.678\,\mu\mathrm{m},40.68\,\mu\mathrm{m}]$.
These values were kept fixed throughout the parameter fitting of section
(\ref{sec:Results}) where $\bar{R}_{\k,a}$ and $\sigma_{\k,a}$
were varied. The discretization then involved dividing these ranges
into $N_{R,\k}$ equal-width volumes or bins, with $N_{R,\k}=20$,
$\k=\n,\p$ for the results shown here. We note that these reduced
size ranges and the subsequent discretization necessitate the lognormal
distributions (\ref{eq:lognormal}), defined on the semi-infinite
range $[0,\infty)$, to be renormalized and the internal parameters
($\mu_{\k,LN}$,$\sigma_{\k,LN}$) to be tweaked so that the distribution
mean and variance are indeed the desired values (e.g., those in Table
\ref{tab:Dimensional-parameters}). This is to ensure that the relation
(\ref{eq:a_k_star}) between $a_{\mathrm{tot},\k}$, $\bar{R}_{\k,a}$
and $\epsilon_{\s,\k}$ is preserved and thus the total active material
volumes remain the experimentally determined values given in Table
\ref{tab:Dimensional-parameters}.

\subsection{Experimental data}

\label{subsec:Experimental-data}

The experimental data that we compare our models to is the validation
data taken from Chen et al. \citep{Chen2020}, consisting of constant
current discharges from 100\% state-of-charge until a cut-off voltage
(2.5 V), followed by a relaxation period of 2 hours where no current
is applied. (We refer to these here as discharge and relaxation experiments.)
They considered experiments at three different C-rates, 0.5C, 1C,
and 1.5C, where, for each C-rate, measurements were taken from three
different cells to give a mean voltage, but also a standard deviation.
All errors relative to experiment are measured relative to this mean
voltage profile, but one standard deviation below and above are also
shown in the figures, given by the thickness of the line. This gives
some context to the size of the errors presented here, in comparison
to the cell-to-cell variation.

\subsection{Parameter fitting}

\label{subsec:Parameter-fitting}

Starting from the parameter set given in Table \ref{tab:Dimensional-parameters},
the models were fitted to the experimental voltage data in a least-squares
sense. For a given model (DFN or MP-DFN) and fitting parameters $\bm{\theta}$
(which depend on the model), the mean-squared error relative to the
experimental data is given by
\begin{equation}
\mathrm{MSE}_{i}(\bm{\theta})=\frac{1}{N_{i}}\sum_{j=1}^{N_{i}}(V(t_{j})-V_{\mathrm{data},j})^{2},
\end{equation}
where $i$ corresponds to the experiment C-rate (0.5C, 1C or 1.5C),
$V_{\mathrm{data},j}$ are the experimental data (at times $t_{j}$),
and $N_{i}$ is the number of data points. The objective or loss function
that we choose to minimize is then the sum of the MSEs across all
C-rates,
\begin{equation}
L(\bm{\theta})=\sum_{i}\mathrm{MSE}_{i}(\bm{\theta})=\mathrm{MSE}_{0.5}(\bm{\theta})+\mathrm{MSE}_{1}(\bm{\theta})+\mathrm{MSE}_{1.5}(\bm{\theta}).\label{eq:L_all_C-rates}
\end{equation}
We employ the mean-squared errors at each C-rate rather than just
the squared error to account for the fact that the data at each C-rate
are of different lengths. A meaningful measure of the error for each
C-rate is given by the root-mean-squared error, $\mathrm{RMSE}_{i}=\sqrt{\mathrm{MSE}_{i}}$,
and one across all C-rates is $\mathrm{RMSE_{total}}=\sqrt{\frac{1}{3}\sum_{i}\mathrm{MSE}_{i}}$,
which are used in Table \ref{tab:Errors}.

The minimization was performed using the Derivative-Free Optimizer
for Least Squares v1.0.2 (DFO-LS) \citep{Cartis2019}. DFO-LS is an
open source robust nonlinear-least-squares minimizer for Python that
does not require derivatives (i.e., the Jacobian) of the objective
with respect to the parameters, and is designed for use with computationally
expensive objective functions. This means it is well-suited to our
purposes, where the evaluation of our objective function (\ref{eq:L_all_C-rates})
requires the simulation of 3 discharge and relaxation experiments,
taking up to 3 minutes in computation time. DFO-LS uses trust region
methods \citep{ConnTrustRegionBook} with the ability to impose bound
constraints, excluding extreme or unphysical regions of the parameter
space from the exploration. The parameter bounds used were conservative,
e.g., with lower bounds of zero, and upper bounds of 5 times the initial
guess (except for the diffusion coefficients, where no upper bounds
were used). Doing this fitting required on the order of 50 function
evaluations for convergence, with absolute and relative tolerances
of $10^{-12}$ and $10^{-20}$, respectively.

\section{Results}

\label{sec:Results}

In this section, we present modelling results that reproduce the experimentally
observed slow relaxation of the voltage after a constant current discharge.
First, the key aspects of the relaxation phenomenon are described,
and then attempts to reproduce these aspects using the DFN model and
the MP-DFN model are made, by fitting a relevant subset of parameters.

\subsection{Modelling the voltage relaxation}

As discussed earlier, there are two key components of the voltage
relaxation for a physical model to capture, which we repeat here:
(1) The final equilibrium voltage after relaxation; (2) The shape
or ``speed'' of the relaxation. We first elucidate the difficulties
in consistently capturing the final voltage, particularly after a
full discharge, and describe one remedy for this scenario, which we
employ here. Then, we move on to the shape of the voltage relaxation. 

\subsubsection{Predicting the equilibrium voltage after relaxation}

\label{subsec:Equi-voltage}

The equilibrium voltage after relaxation in our physical models is
given simply by the OCV, $V_{\text{eq}}=U_{\p}(\tilde{c}_{\s,\p,\text{eq}})-U_{\n}(\tilde{c}_{\s,\n,\text{eq}})$,
where $\tilde{c}_{\s,\p,\text{eq}}$ and $\tilde{c}_{\s,\n,\text{eq}}$
are the final uniform stoichiometries in the positive and negative
electrodes. Thus, predicting $V_{\text{eq}}$ amounts to predicting
$\tilde{c}_{\s,\p,\text{eq}}$ and $\tilde{c}_{\s,\n,\text{eq}}$.
However, $V_{\text{eq}}$ can depend very sensitively on these stoichiometries,
particularly, close to 0\% and 100\% state-of-charge, where the OCPs,
$U_{\p}$ and $U_{\n}$, are singular with large gradients. Therefore,
small errors in the model's internal states can result in significant
errors in $V_{\text{eq}}$. This is not an issue if the experiment
undertaken is a (dis)charge for a predefined length of time, when
the final model stoichiometries are determined straightforwardly and
accurrately via charge and lithium conservation, the only error being
that made during electrode balancing. It is an issue, however, if
the (dis)charge stopping criterion is a cut-off voltage close to 0\%
or 100\% state-of-charge. This is evident in the model validations
of, e.g., \citep{Ecker2015a,Ecker2015b,Schmalstieg2018a,Schmalstieg2018b,Sulzer2019b,Chen2020},
which include validation experiments of this type. 

\begin{figure}
\includegraphics[width=1\textwidth]{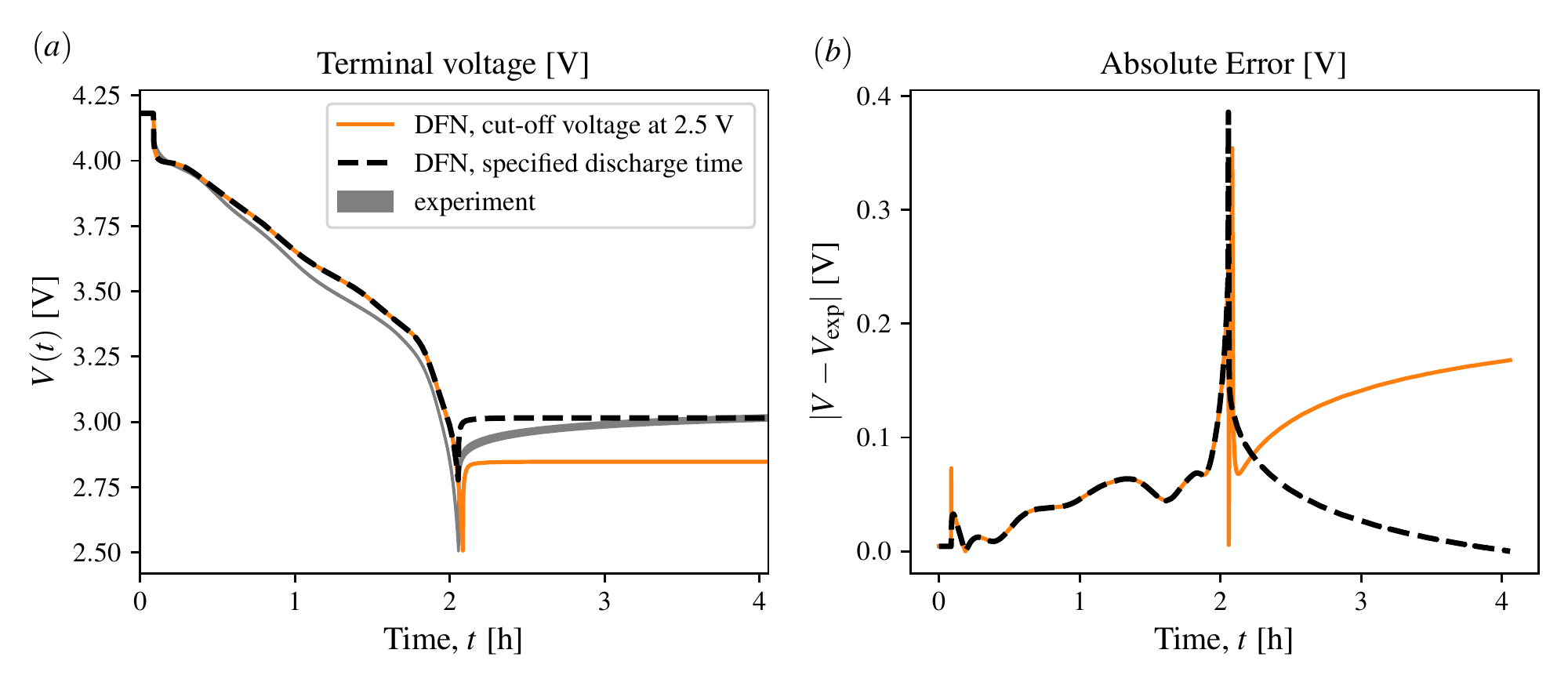}

\caption{\label{fig:DFN_cutoff_voltage}A discharge (0.5C) and relaxation,
comparing two different DFN model simulations (section \ref{subsec:DFN-model},
parameters in Table \ref{tab:Dimensional-parameters}) to experimental
data of \citep{Chen2020}. One simulation discharges until the cut-off
voltage, 2.5 V, and the other for a specified time of 7084.80 s, matching
the experimental time (to within approximately 5 s). Panel ($a$)
shows the voltage, and ($b$) the absolute error of the simulations
relative to the experiments.}
\end{figure}

The experimental data which we consider here (see section \ref{subsec:Experimental-data})
consist of constant current discharges from a fully-charged cell until
the lower cut-off voltage of 2.5 V is reached, followed by 2 hours
of relaxation. Fig. \ref{fig:DFN_cutoff_voltage} shows the experimental
data for a 0.5C discharge, and two numerical simulations of the DFN
model, consisting of a 0.5C discharge: (i) until the cut-off voltage
(2.5 V); (ii) for a specified amount of time, chosen to match the
discharge time observed in the experiment. The DFN model is identical
to the one in \citep{Chen2020}, which was validated by this data,
and hence reasonably reproduces the discharge portion. If the cut-off
voltage criterion is used, the model reaches this voltage slightly
later ($\sim102$ s over a 2 hour discharge) compared to the experiment,
resulting in a significant error in the final rest voltage. (In Chen
et al. \citep{Chen2020}, the diffusion coefficient in the negative
electrode was changed for each C-rate to alter the discharge time
and hence reproduce the equilibrium voltage.) However, the simulation
with the specified discharge time, despite not reaching the cut-off
voltage, captures the final rest voltage excellently. In fact, the
experimental data (which is processed from raw data) was provided
at approximately 5 second intervals, and hence the time until cut-off
could only be extracted up to an error of $\pm5$ seconds. The time
was then tweaked within this window to exactly fit the model to the
final equilibrium voltage. Using the DFN, the discharge times for
the 0.5C, 1C, 1.5C experiments were found to be 7084.80 s, 3544.56
s, 2360.23 s, respectively.

Even though the time-specified simulation does not reach the cut-off
voltage, its peak error (see Fig. \ref{fig:DFN_cutoff_voltage}($b$)),
is comparable to the peak error of the simulation that does reach
it. Furthermore, the times specified above are not just applicable
to the DFN but also the MP-DFN, which we demonstrate in section \ref{subsec:Comparison_fitting}.
Therefore, for the fitting of the voltage relaxation shape in section
\ref{subsec:Shape_of_voltage} and \ref{subsec:Comparison_fitting},
we fix the discharge times to be those above, ensuring the equilibrium
voltage is not affected by the variation of the diffusion coefficients
and electrode microstructure parameters. As we will see, the voltage
at cut-off moves closer to 2.5 V naturally as the models are fit to
the rest of the voltage curve.

\subsubsection{Shape of the voltage relaxation}

\label{subsec:Shape_of_voltage}

As observed in Fig. \ref{fig:DFN_cutoff_voltage}, and also other
relaxation experiments in the literature \citep{Ecker2015b,Schmalstieg2018b,Sulzer2019b},
the voltage relaxes to the equilibrium value monotonically, but typically
slower than seen in simulations. This implies an internal heterogeneity
which equilibriates on a very long timescale (on the order of several
hours) which does not seem to be present in the standard DFN-type
model. The slowest timescale candidates in the DFN that may be responsible
for this slow relaxation are the diffusion timescale of lithium in
the electrode particles, and the reaction timescales, given by
\begin{align}
\tau_{\k} & =\frac{(\bar{R}_{\k,a})^{2}}{D_{\s,\k}},\qquad\k=\n,\p,\label{eq:diff_timescale}\\
\tau_{\mathrm{r},\k} & =\frac{F}{m_{\k}a_{\mathrm{tot},\k}c_{\e,\mathrm{typ}}^{1/2}}=\frac{F\bar{R}_{\k,a}}{3m_{\k}\epsilon_{\s,\k}c_{\e,\mathrm{typ}}^{1/2}},\qquad\k=\n,\p.\label{eq:reaction_timescale}
\end{align}
These are timescales representative of the entire electrode(s). A
model with a single particle size per electrode, such as the DFN,
thus has a single diffusion timescale in each. For models with many
particle sizes (such as the MP-DFN), there is a diffusion timescale
for each particle size, with those larger than the mean radius $\bar{R}_{\k,a}$
having longer diffusion timescales, and therefore relaxations, than
the mean particle. This distribution of diffusion timescales is one
method of realistically including longer timescales in a physical
model of the battery. We may vary the overall diffusion and reaction
timescales (\ref{eq:diff_timescale})-(\ref{eq:reaction_timescale})
by varying $D_{\s,\k}$ and $\bar{R}_{\k,a}$, and control the spread
of the distribution of timescales via the standard deviations $\sigma_{a,\k}$.
In the next section we proceed to vary these microscale parameters
in the DFN and MP-DFN models with emphasis on how they might influence
reproducing the slow relaxation in the experiments.

\subsection{Comparison of models and experiment}

\label{subsec:Comparison_fitting}

In this section, we directly compare the DFN and MP-DFN models to
the experimental discharge and relaxation data (see section \ref{subsec:Experimental-data})
across a range of C-rates. By varying the microscale parameters $D_{\s,\k}$,
$\bar{R}_{\k,a}$, $\sigma_{a,\k}$, $\k=\n,\p$, best fits of the
DFN and MP-DFN to the voltage profiles are presented and analyzed.
The relevant parameter values, before and after fitting, are given
in Table \ref{tab:Fitted_params}.

\subsubsection{DFN model}

First, to demonstrate that the DFN model (with a single particle size
per electrode) cannot reproduce the slow relaxation phenomenon, we
attempt a best fit to the experimental data, varying the four parameters
$D_{\s,\k}$, $\bar{R}_{\k,a}$, $\k=\n,\p$. Although the parameter
set in Table \ref{tab:Dimensional-parameters} is equivalent to the
set in Chen et al. \citep{Chen2020}, and several parameters were
already tuned by them to fit the DFN to this data, their fitting was
done by hand using trial and error. Also, their diffusion coefficient
in the negative electrode, $D_{\s,\n}$, was made to depend on the
C-rate in order to fit the final rest voltages. However, we take $D_{\s,\n}$
to be independent of C-rate, and ensure the final rest voltages are
accurate by fixing the cut-off times\textemdash see section \ref{subsec:Equi-voltage}.
Therefore, to eliminate the possibility that the DFN can fit the slow
relaxation with a better choice of parameter values, the fitting is
performed here for all C-rate experiments simultaneously using a numerical
optimization package\textemdash see section \ref{subsec:Parameter-fitting}. 

\begin{figure}
\includegraphics[width=1\textwidth]{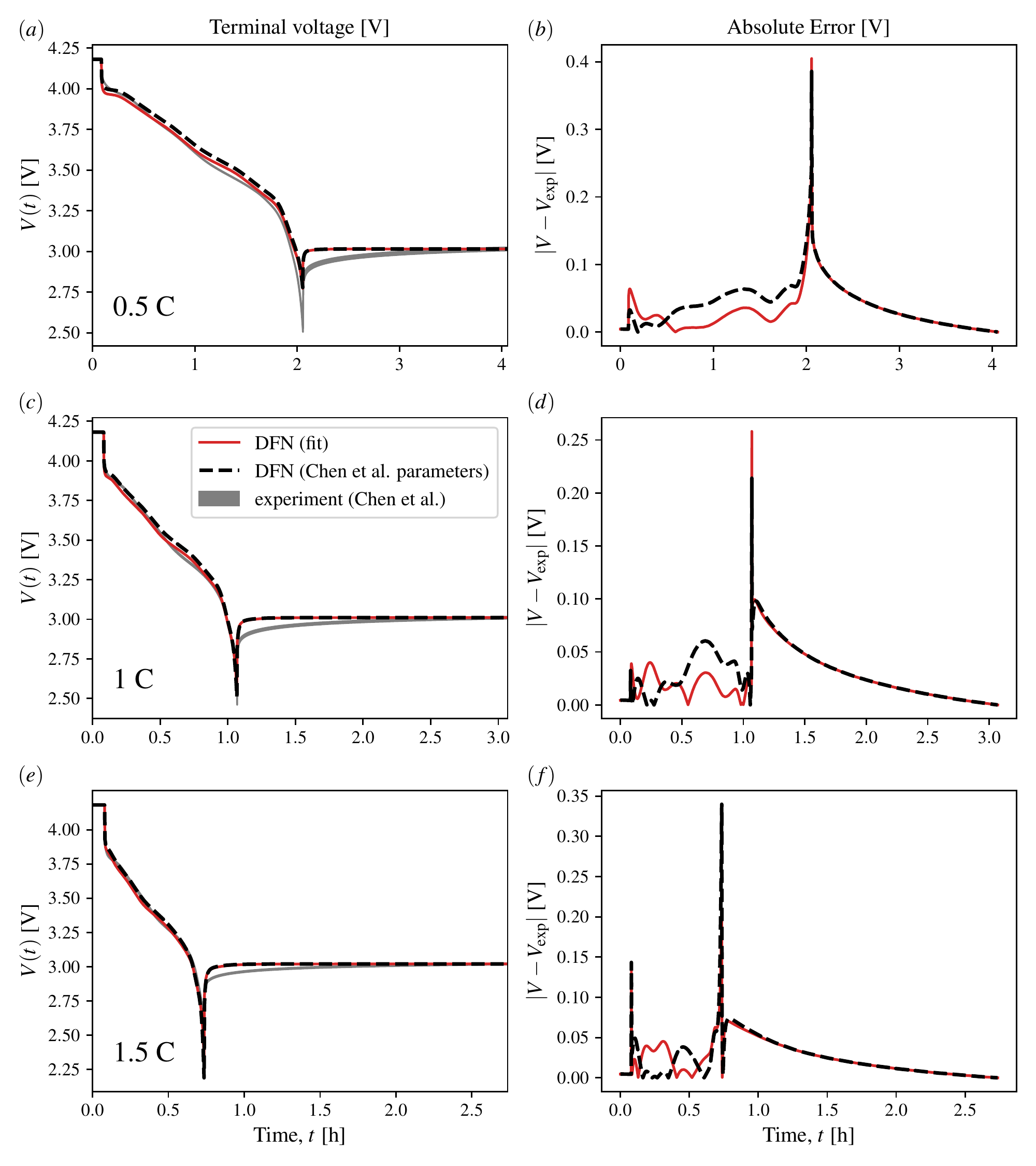}

\caption{\label{fig:DFN_fitting}Comparison between discharge and relaxation
experiments \citep{Chen2020} and simulations of the DFN model (section
\ref{subsec:DFN-model}), for parameter set from \citep{Chen2020}
(Table \ref{tab:Dimensional-parameters}), and fitted parameters (Table
\ref{tab:Fitted_params}). The rows correspond to different C-rates
(0.5C, 1C, 1.5C), with the terminal voltage shown in the left column,
and the error relative to experiment shown on the right. Width of
experimental curve is 2 standard deviations (centred on the mean)
measured from three cells (see section \ref{subsec:Experimental-data}
or \citep{Chen2020}).}
\end{figure}

Fig. \ref{fig:DFN_fitting} shows the DFN results for the three discharge
and relaxation experiments (0.5C, 1C, 1.5C), for the Chen et al. parameter
set in Table \ref{tab:Dimensional-parameters}, and after fitting.
The voltages and the absolute error relative to the experiments are
shown over time for each C-rate. Each simulation captures the final
equilibrium voltage after relaxation since the discharge times have
been specified to match these experimental data. Relative to the parameters
from \citep{Chen2020}, $D_{\s,\n}$ and $\bar{R}_{\n,a}$ have been
increased ($\tau_{\n}$ decreased) in the fitted solution, and $D_{\s,\p}$
and $\bar{R}_{\p,a}$ have been decreased ($\tau_{\p}$ increased).
The fitting results in a modest improvement over the discharge portions
for all C-rates, with $\mathrm{RMSE_{total}}$ reduced from 42.2 mV
to 37.1 mV\textemdash see Table \ref{tab:Errors}. However, the relaxation
portion is almost unchanged, with the speed of relaxation still greatly
overestimated. Numerous different fits were attempted, employing different
random seeds in the optimization and various parameter bounds, but
no further improvement could be attained. It may be possible to reduce
the relaxation speed (and hence the error) of the DFN by fitting to
only the relaxation portions but, as these results demonstrate, it
does not appear possible while constrained to simultaneously fit the
discharge portions.

\subsubsection{MP-DFN model}

\begin{figure}
\includegraphics[width=1\textwidth]{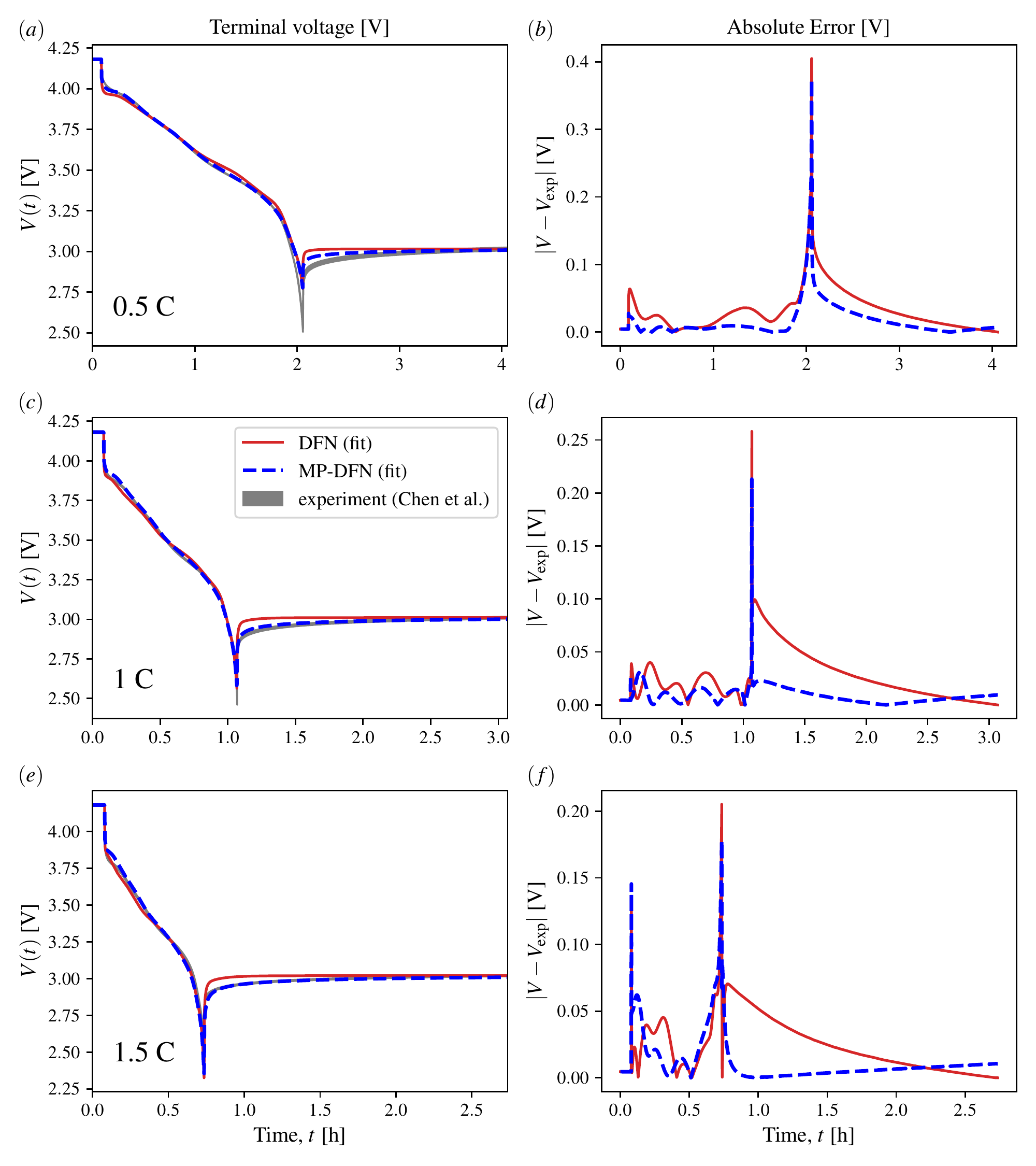}

\caption{\label{fig:MPDFN_fitting}Comparison between discharge and relaxation
experiments \citep{Chen2020} and fitted simulations of the MP-DFN
model (section \ref{sec:Modelling}). Also shown is the fitted DFN
model, with fitted parameters in Table \ref{tab:Fitted_params}. The
rows correspond to different C-rates (0.5C, 1C, 1.5C), with the terminal
voltage shown in the left column, and the error relative to experiment
shown on the right. Width of experimental curve is 2 standard deviations
(centred on the mean) measured from three cells (see section \ref{subsec:Experimental-data}
or \citep{Chen2020}).}
\end{figure}

\begin{figure}
\begin{centering}
\includegraphics[width=1\textwidth]{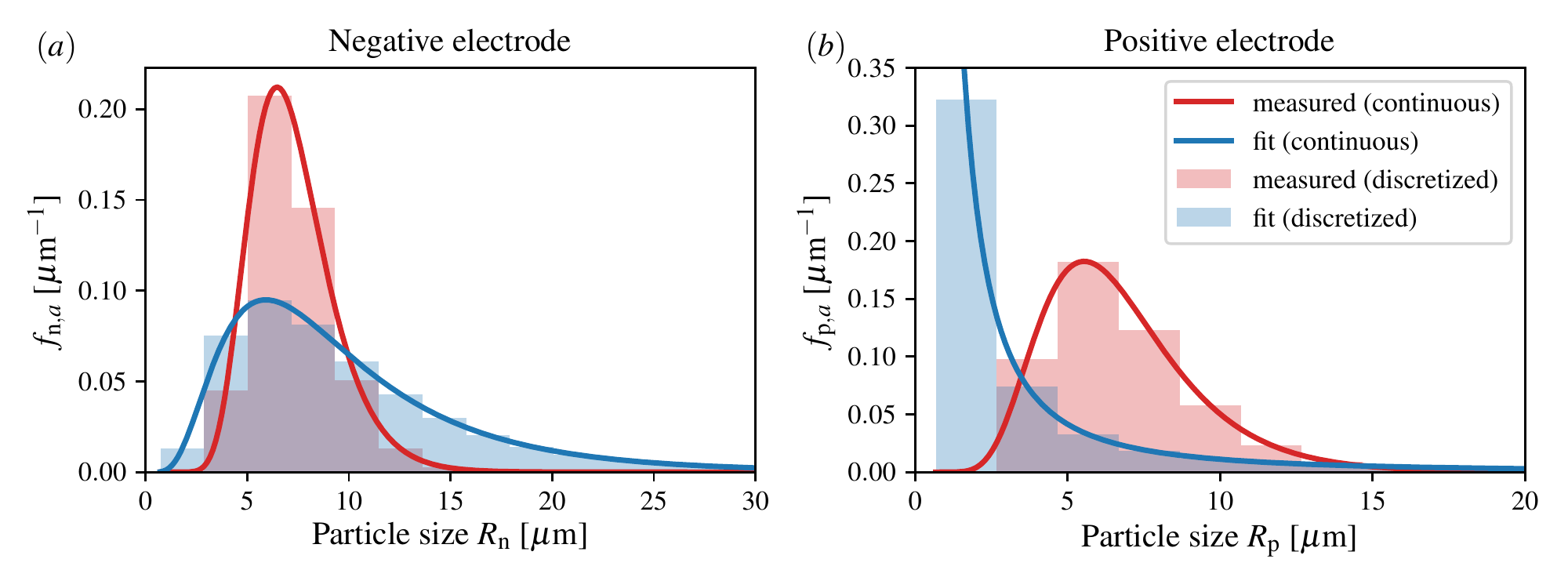}
\par\end{centering}
\caption{\label{fig:Final_fitted_PSDs}Area-weighted particle-size distributions
$f_{\protect\k,a}(R_{\protect\k})$, $\protect\k=\protect\n,\protect\p$
from direct measurements \citep{Chen2020} (red) and from fitting
the MP-DFN model to voltage relaxation data (blue). The continuous
lognormal densities are shown, as well as the discrete distributions
that were employed in the PyBaMM simulations. For the means ($\bar{R}_{\protect\k,a}$)
and standard deviations ($\sigma_{a,\protect\k}$) see Table \ref{tab:Fitted_params}.}
\end{figure}

We now present the results of the MP-DFN model compared to the experimental
data, including the best fit under the variation of the six parameters
$D_{\s,\k}$, $\bar{R}_{\k,a}$, $\sigma_{a,\k}$, $\k=\n,\p$. The
same numerical optimization methods were used as the DFN, but now
there are six fitting parameters rather than four. Fig. \ref{fig:MPDFN_fitting}
shows the fitted MP-DFN results for the three discharge and relaxation
experiments (0.5C, 1C, 1.5C),  but also the best fit of the DFN for
comparison.  The  MP-DFN shows an almost uniform reduction in error
across all times and C-rates, compared to the best-fit DFN. Relative
to the measured parameters (Table \ref{tab:Fitted_params}), $D_{\s,\n}$
and $\bar{R}_{\n,a}$ are increased ($\tau_{\n}$ decreased) and $D_{\s,\p}$
and $\bar{R}_{\p,a}$ are decreased ($\tau_{\p}$ increased), but
the spreads of both particle-size distributions, $\sigma_{a,\n}$
and $\sigma_{a,\p}$, are increased\textemdash the fitted distributions
can be seen in Fig. \ref{fig:Final_fitted_PSDs}. The error during
discharge portions is lower for the MP-DFN but, crucially, the relaxation
has been slowed down to produce excellent fits to the experimental
relaxations. As a result, the RMSE is reduced to 29.8 mV (0.5C), 12.3
mV (1C), 21.7 mV (1.5C), 22.4 mV (all C-rates), an approximately 40\%
overall error reduction relative to the best DFN fit.

The poor fit near the current cut-off for 0.5C is due to a slow voltage
drop at the end of discharge. The fit can be improved considerably
if we modify one parameter, $D_{\s,\n}$ say, for this C-rate only.
If this is done, the best-fit value for $D_{\s,\n}$ is $1.30\times10^{-13}$
{\footnotesize{}$\text{m}\ensuremath{^{2}}\text{s}\ensuremath{^{-1}}$}
and the agreement with experiment becomes remarkable, as shown in
Fig. \ref{fig:MP-DFN_fitting_05C_only}, with an RMSE of 10.3 mV.
We remark that in the original parametrisation of Chen et al. \citep{Chen2020},
$D_{\s,\n}$ had to be changed for each C-rate to correctly predict
the rest voltage. (The fits in Fig. \ref{fig:MP-DFN_fitting_05C_only}
are for $D_{\s,\n}$ constant and independent of C-rate, with prediction
of the rest voltage ensured by fixing the cut-off times.) Indeed,
the negative electrode of these cells is actually composed of two
active materials (silicon in addition to graphite), and $D_{\s,\n}$
is only an effective diffusion coefficient accounting for both in
a single phase. With this in mind, it is remarkable that we could
find such good agreement to the experiments. 

\begin{figure}
\begin{centering}
\includegraphics[width=0.6\textwidth]{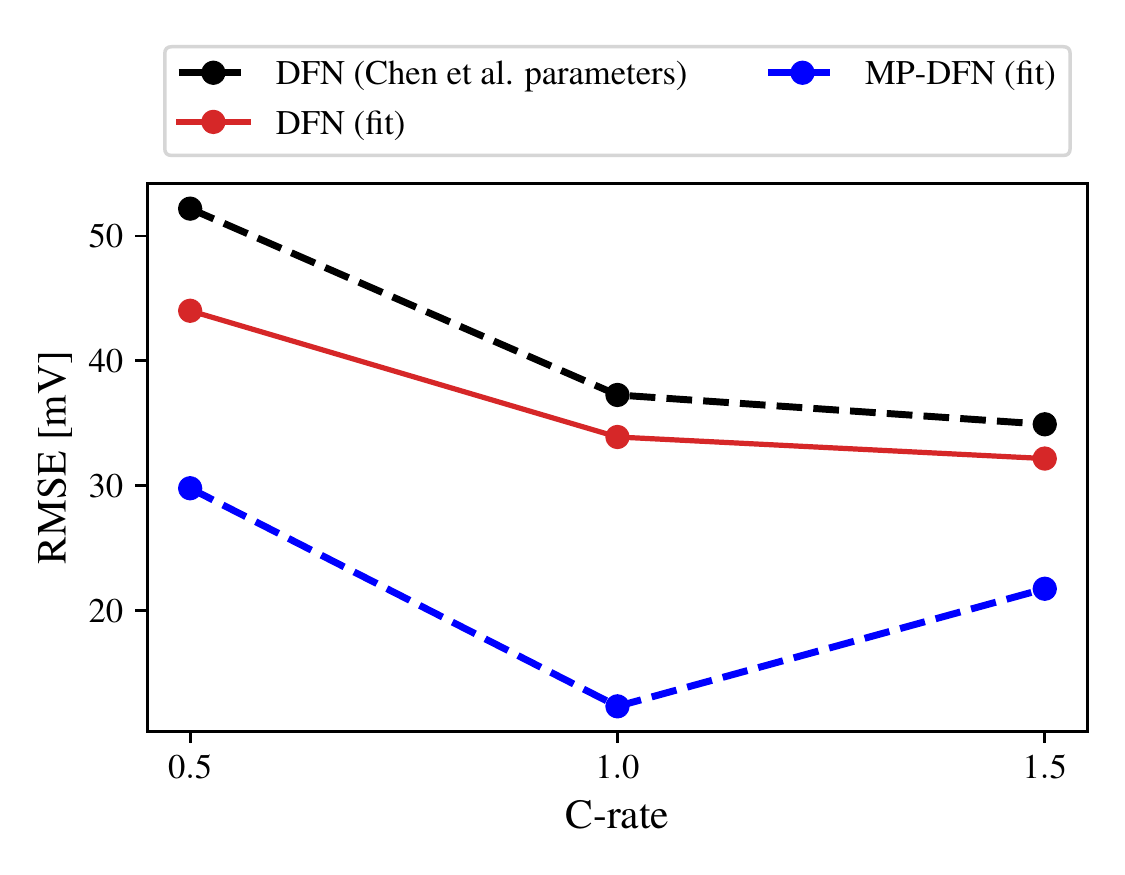}
\par\end{centering}
\caption{\label{fig:RMSE_values}Root-mean-squared errors (RMSE) of the DFN
and MP-DFN models relative to experimental data \citep{Chen2020},
for parameters from \citep{Chen2020} (Table \ref{tab:Dimensional-parameters})
and fitted parameters (Table \ref{tab:Fitted_params}). Errors are
shown for the three different C-rates: 0.5C, 1C, 1.5C. }
\end{figure}

\begin{figure}
\begin{centering}
\includegraphics[width=1\textwidth]{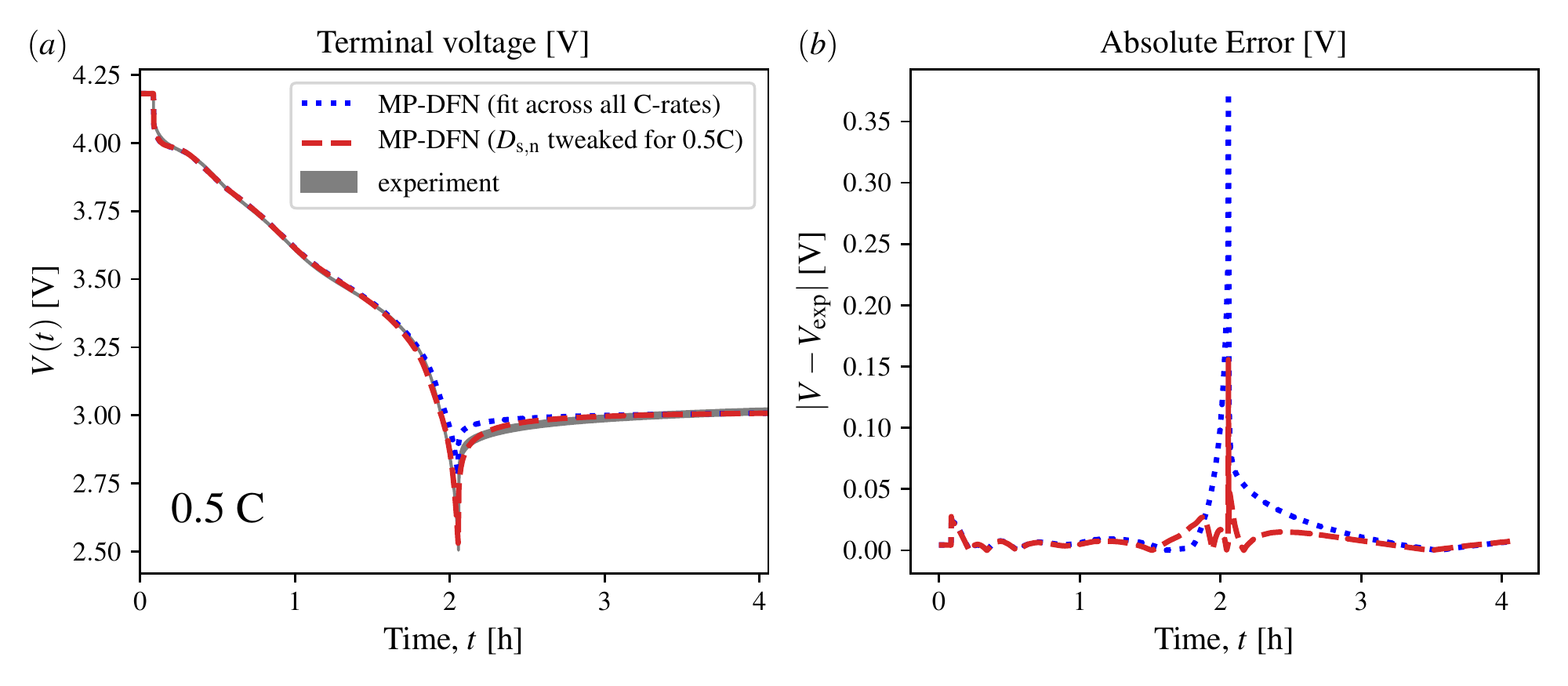}
\par\end{centering}
\caption{\label{fig:MP-DFN_fitting_05C_only}MP-DFN, with parameters fitted
across all C-rates, and when $D_{\protect\s,\protect\n}$ is then
tweaked for 0.5C only. Dotted line is for parameters in Table \ref{tab:Fitted_params},
and dashed line is the same but with $D_{\protect\s,\protect\n}$
changed to $1.30\times10^{-13}$ {\footnotesize{}$\text{m}\ensuremath{^{2}}\text{s}\ensuremath{^{-1}}$}
, giving RMSE of 10.3 mV, and cut-off voltage 2.531 V.}
\end{figure}

The errors (RMSE) of each model (DFN and MP-DFN), for the parameters
from \citep{Chen2020} and our fitted parameters, are summarized in
Fig. \ref{fig:RMSE_values} and Table \ref{tab:Errors}. The MP-DFN
model, using either parameter set, exhibits smaller errors than the
DFN (except for one case at 1.5C), and the fit MP-DFN model has the
lowest error for any C-rate.

Finally, we comment on the differences between the fitted size distributions
and those measured directly in \citep{Chen2020}\textemdash both are
shown in Fig. \ref{fig:Final_fitted_PSDs}. Since the measurements
are done on cross-sections of the electrodes (and therefore the active
particles), one may expect the measured mean radii to underestimate
the true mean, and indeed, the fitted mean for the negative electrode
is larger than the measured value. However, the difference in the
mean for the positive electrode is more significant. The fitted mean
is 40\% lower than the measured value which, together with the larger
variance, results in significantly more smaller particles, i.e., those
around 1 $\mu$m or less. One explanation is that NMC811, the positive
electrode active material, is known to form large secondary particle
agglomerates ($\sim$1-15 $\mu$m) out of many smaller primary particles
($<1\,\mu$m) \citep{Heenan2020}. From their SEM images, the particles
measured by \citep{Chen2020} may be the secondary agglomerates, based
on their size. If so, our results suggest that the primary particle
size should be used in physical models rather than the secondary one.

\begin{table}
\begin{centering}
\begin{tabular}{clcccc}
\hline 
\multirow{2}{*}{{\footnotesize{}\makecell[l]{Dimensional \\ parameter}}} & \multirow{2}{*}{{\footnotesize{}unit}} & \multirow{2}{*}{{\footnotesize{}\makecell[l]{Electrode \\ $\k\in \{\n,\p\}$}}} & {\footnotesize{}From \citep{Chen2020}} & \multicolumn{2}{c}{{\footnotesize{}Fit}}\tabularnewline
\cline{4-6} 
 &  &  & {\footnotesize{}DFN/MP-DFN} & {\footnotesize{}DFN} & {\footnotesize{}MP-DFN}\tabularnewline
\hline 
\multirow{2}{*}{{\footnotesize{}$D_{\s,\k}$}} & \multirow{2}{*}{{\footnotesize{}$\text{m}\ensuremath{^{2}}\text{s}\ensuremath{^{-1}}$}} & {\footnotesize{}n} & {\footnotesize{}$5.10\times10^{-14}$} & {\footnotesize{}$2.70\times10^{-13}$} & {\footnotesize{}$5.25\times10^{-13}$}\tabularnewline
 &  & {\footnotesize{}p} & {\footnotesize{}$6.75\times10^{-15}$} & {\footnotesize{}$1.49\times10^{-15}$} & {\footnotesize{}$1.76\times10^{-14}$}\tabularnewline
\multirow{2}{*}{{\footnotesize{}$\bar{R}_{\mathrm{\k},a}$}} & \multirow{2}{*}{{\footnotesize{}m}} & {\footnotesize{}n} & {\footnotesize{}$7.28\times10^{-6}$} & {\footnotesize{}$14.6\times10^{-6}$} & {\footnotesize{}$9.98\times10^{-6}$}\tabularnewline
 &  & {\footnotesize{}p} & {\footnotesize{}$6.78\times10^{-6}$} & {\footnotesize{}$3.39\times10^{-6}$} & {\footnotesize{}$4.10\times10^{-6}$}\tabularnewline
\multirow{2}{*}{{\footnotesize{}$\sigma_{a,\k}$}} & \multirow{2}{*}{{\footnotesize{}m}} & {\footnotesize{}n} & {\footnotesize{}$2.08\times10^{-6}$} & {\footnotesize{}-} & {\footnotesize{}$6.15\times10^{-6}$}\tabularnewline
 &  & {\footnotesize{}p} & {\footnotesize{}$2.59\times10^{-6}$} & {\footnotesize{}-} & {\footnotesize{}$5.43\times10^{-6}$}\tabularnewline
\multirow{2}{*}{{\footnotesize{}$a_{\mathrm{tot},\mathrm{\k}}$}} & \multirow{2}{*}{{\footnotesize{}$\text{m}^{-1}$}} & {\footnotesize{}n} & {\footnotesize{}$3.09\times10^{5}$} & {\footnotesize{}$1.55\times10^{5}$} & {\footnotesize{}$2.26\times10^{5}$}\tabularnewline
 &  & {\footnotesize{}p} & {\footnotesize{}$2.94\times10^{5}$} & {\footnotesize{}$5.88\times10^{5}$} & {\footnotesize{}$4.87\times10^{5}$}\tabularnewline
\hline 
\end{tabular}
\par\end{centering}
\caption{\label{tab:Fitted_params}Parameters from \citep{Chen2020} (Table
\ref{tab:Dimensional-parameters}) and fitted parameter values corresponding
to the fits of the DFN and MP-DFN models to the discharge and relaxation
experimental data \citep{Chen2020}.}
\end{table}

\begin{table}
\begin{centering}
\begin{tabular}{clcccc}
\hline 
\multirow{3}{*}{{\footnotesize{}Quantity {[}unit{]}}} & \multirow{3}{*}{{\footnotesize{}Experiment}} & \multicolumn{4}{c}{{\footnotesize{}Parameter set}}\tabularnewline
\cline{3-6} 
 &  & \multicolumn{2}{c}{{\footnotesize{}From \citep{Chen2020}}} & \multicolumn{2}{c}{{\footnotesize{}Fit}}\tabularnewline
\cline{3-6} 
 &  & {\footnotesize{}DFN} & {\footnotesize{}MP-DFN} & {\footnotesize{}DFN} & {\footnotesize{}MP-DFN}\tabularnewline
\hline 
\multirow{4}{*}{{\footnotesize{}Voltage RMSE {[}mV{]}}} & {\footnotesize{}0.5C} & {\footnotesize{}\cellcolor[hsb]{0,0.95,1}52.2} & {\footnotesize{}\cellcolor[hsb]{0,0.69,1}37.8} & {\footnotesize{}\cellcolor[hsb]{0,0.80,1}44.0} & {\footnotesize{}\cellcolor[hsb]{0,0.54,1}29.8}\tabularnewline
 & {\footnotesize{}1C} & {\footnotesize{}\cellcolor[hsb]{0,0.68,1}37.3} & {\footnotesize{}\cellcolor[hsb]{0,0.46,1}25.2} & {\footnotesize{}\cellcolor[hsb]{0,0.62,1}33.9} & {\footnotesize{}\cellcolor[hsb]{0,0.22,1}12.3}\tabularnewline
 & {\footnotesize{}1.5C} & {\footnotesize{}\cellcolor[hsb]{0,0.64,1}34.9} & {\footnotesize{}\cellcolor[hsb]{0,1,1}54.7} & {\footnotesize{}\cellcolor[hsb]{0,0.59,1}32.2} & {\footnotesize{}\cellcolor[hsb]{0,0.39,1}21.7}\tabularnewline
 & {\footnotesize{}All C-rates} & {\footnotesize{}\cellcolor[hsb]{0,0.77,1}42.2} & {\footnotesize{}\cellcolor[hsb]{0,0.75,1}41.1} & {\footnotesize{}\cellcolor[hsb]{0,0.68,1}37.1} & {\footnotesize{}\cellcolor[hsb]{0,0.41,1}22.4}\tabularnewline
\hline 
\multirow{3}{*}{{\footnotesize{}Voltage at cut-off {[}V{]}}} & {\footnotesize{}0.5C} & {\footnotesize{}2.778} & {\footnotesize{}2.759} & {\footnotesize{}2.773 } & {\footnotesize{}2.777}\tabularnewline
 & {\footnotesize{}1C} & {\footnotesize{}2.513} & {\footnotesize{}2.457} & {\footnotesize{}2.564} & {\footnotesize{}2.582}\tabularnewline
 & {\footnotesize{}1.5C} & {\footnotesize{}2.188} & {\footnotesize{}2.059} & {\footnotesize{}2.325} & {\footnotesize{}2.354}\tabularnewline
\hline 
\end{tabular}
\par\end{centering}
\caption{\label{tab:Errors}Root-mean-squared errors (RMSE) of the DFN and
MP-DFN models relative to experimental data \citep{Chen2020}, using
parameters from \citep{Chen2020} (see Table \ref{tab:Dimensional-parameters})
and fitted parameters (Table \ref{tab:Fitted_params}). Also shown
are the voltages just before current cut-off. Errors are shown for
the three different C-rates: 0.5C, 1C, 1.5C, and RMSE across all C-rates
calculated as in section \ref{subsec:Parameter-fitting}. If $D_{\protect\s,\protect\n}$
is changed to $1.30\times10^{-13}\text{m}\ensuremath{^{2}}\text{s}\ensuremath{^{-1}}$
for 0.5C, MP-DFN (fit), then RMSE = 10.3 mV (0.5C), 15.6 mV (all C-rates). }
\end{table}

\section{Conclusions and Future Work}

In this paper we explored, using physical models, the voltage relaxation
phenomenon previously observed but not physically explained by several
parametrization and modelling studies on lithium ion batteries. We
focused on the discharge and relaxation experimental data of Chen
et al. \citep{Chen2020}, taken from a set of commercial cells (LGM50,
cylindrical 21700), wherein the relaxation after a full discharge
was inadequately reproduced by the physical model they considered:
the Doyle\textendash Fuller\textendash Newman (DFN) model that is
used extensively in the literature. This poor fitting of the DFN model
is seen across numerous other parametrization studies.

Here we considered an extension of the DFN model to include a distribution
of particle sizes in the active material of each electrode, rather
than a single size as assumed in the DFN. This model, denoted the
Many-Particle-DFN (MP-DFN), was initially parametrized by modifying
the DFN parameter set in \citep{Chen2020} to include lognormal (area-weighted)
particle-size distributions (PSDs) fitted to measurements taken (but
not used) in \citep{Chen2020}. This process of adapting an existing
DFN parameter set for use with the MP-DFN is detailed, with attention
given to numerical robustness and the use of physically relevant mean
particle radii, in order to facilitate the process for other cells
and chemistries.

The discharge and relaxation simulations of the MP-DFN model were
then compared to the experimental data, and contrasted with those
of the DFN model. To account for experimental uncertainty in the electrode
microstructure properties, the PSDs (i.e., the mean radii and standard
deviations) and effective diffusivity of lithium in the active material
were also optimized to best fit the voltage data across the full range
of C-rates (0.5C, 1C, 1.5C). The numerical simulations were performed
using the flexible and robust software package PyBaMM, and the parameter
optimization used a derivative-free nonlinear least squares minimizer,
DFO-LS, written in Python. 

The final equilibrium voltage after the relaxation period was able
to be consistently captured, by both the MP-DFN and DFN, by matching
the discharge time to that observed in each experiment. We found that
the DFN always overestimates the speed of the voltage relaxation,
even with parameter optimization, showing that a timescale of sufficient
length is likely not possible within the model. However, our results
show that the MP-DFN can exhibit a relaxation slow enough to match
that of the experiments, owing to a distribution of diffusion timescales
and greater scope for internal heterogeneities. Even for the experimentally
measured microstructure parameters, the MP-DFN better matched the
experiments during discharge as well as the relaxation. This can be
greatly improved with parameter optimization, where the spread of
the PSDs (and hence distribution of timescales) are increased, resulting
in remarkable agreement for all C-rates, and an average error of 22.4
mV, or 15.6 mV if the diffusivity in the negative electrode is modified
for one value of the C-rate (0.5C).

Avenues for future work include the consideration of lithium ion batteries
with different electrode chemistries. The cells modelled here have
negative electrodes that are a composite of two active materials (graphite
and silicon), but only one material was modelled for simplicity. Experiments
on cells without composite electrodes, or models extended to account
for the composite explicitly, may allow even better agreement without
necessitating that, e.g., diffusivities depend on the C-rate. One
could also consider (dis)charge relaxation experiments with cut-off
chosen at other states of charge, far from 0\% or 100\% where the
OCPs are steep, reducing the difficulties in capturing the equilibrium
states. Finally, the MP-DFN is computationally expensive, which motivates
deriving reduced-order models that can display the same behaviour\textemdash this
is currently under way.

\subsubsection*{Acknowledgments}

The authors acknowledge funding provided by The Faraday Institution,
grant number \\
EP/S003053/1, FIRG003. We also thank the Energy Materials Group (University of Birmingham) and the Energy Group (University of Warwick)
for providing the experimental data from Chen et al. \citep{Chen2020}.

\appendix

\section{MP-DFN model: dimensionless governing equations}

In this appendix we state the dimensionless variables and equations
that are solved numerically in PyBaMM. We employ a dimensionless scheme
similar to that of Marquis et al. \citep{Marquis2019}, summarised
in Table \ref{tab:Dimensionless-variables}. Dimensional quantities
are now indicated by an asterisk in this appendix to distinguish them
from their dimensionless counterparts. The physically relevant timescales
and resulting dimensionless parameters are given in Table \ref{tab:Dimensionless-parameters}.
The dimensionless problem is summarised below.

\begin{table*}
\begin{centering}
\begin{tabular}{cllccc}
\hline 
\multirow{2}{*}{{\scriptsize{}Parameter}} & \multirow{2}{*}{{\scriptsize{}Definition}} & \multirow{2}{*}{{\scriptsize{}Interpretation}} & \multicolumn{3}{c}{{\scriptsize{}Region $\k$}}\tabularnewline
 &  &  & {\scriptsize{}$\n$} & {\scriptsize{}$\sep$} & {\scriptsize{}$\p$}\tabularnewline
\hline 
\hline 
{\scriptsize{}$\tau_{\mathrm{d}}^{*}$} & {\scriptsize{}$F^{*}c_{\n,\mathrm{max}}^{*}L^{*}/i_{\mathrm{typ}}^{*}$} & {\scriptsize{}Discharge timescale} & \multicolumn{3}{c}{{\scriptsize{}$1.10\times10^{5}/\mathcal{C}$}}\tabularnewline
{\scriptsize{}$\tau_{\e}^{*}$} & {\scriptsize{}$L^{*2}/D_{\e,\mathrm{typ}}^{*}$} & {\scriptsize{}Diffusion timescale in the electrolyte} & \multicolumn{3}{c}{{\scriptsize{}$1.69\times10^{2}$}}\tabularnewline
{\scriptsize{}$\tau_{\k}^{*}$} & {\scriptsize{}$(\bar{R}_{\k,a}^{*})^{2}/D_{\s,\k}^{*}$} & {\scriptsize{}\makecell[l]{ Diffusion timescale in the average-sized \\ electrode particle}} & {\scriptsize{}$1.04\times10^{3}$} & {\scriptsize{}-} & {\scriptsize{}$6.81\times10^{3}$}\tabularnewline
{\scriptsize{}$\tau_{\mathrm{r},\k}^{*}$} & {\scriptsize{}$F^{*}/(m_{\k}^{*}a_{\mathrm{tot},\k}^{*}(c_{\e,0}^{*})^{1/2})$} & {\scriptsize{}Reaction timescale in the electrode} & {\scriptsize{}$1.23\times10^{4}$} & {\scriptsize{}-} & {\scriptsize{}$2.34\times10^{3}$}\tabularnewline
\hline 
{\scriptsize{}$\mathcal{C}$} & {\scriptsize{}$i_{\mathrm{typ}}^{*}/C^{*}$} & {\scriptsize{}C-rate} & \multicolumn{3}{c}{{\scriptsize{}0.5, 1 or 1.5}}\tabularnewline
{\scriptsize{}$\mathcal{C}_{\k}$} & {\scriptsize{}$\tau_{\k}^{*}/\tau_{\mathrm{d}}^{*}$} & {\scriptsize{}\makecell[l]{Ratio of solid diffusion \\ to discharge timescales} } & {\scriptsize{}$9.41\times10^{-3}\mathcal{C}$} & {\scriptsize{}-} & {\scriptsize{}$6.16\times10^{-2}\mathcal{C}$}\tabularnewline
{\scriptsize{}$\mathcal{C}_{\mathrm{r},\k}$} & {\scriptsize{}$\tau_{\mathrm{r},\k}^{*}/\tau_{\mathrm{d}}^{*}$} & {\scriptsize{}\makecell[l]{Ratio of reaction to  \\ discharge timescales} } & {\scriptsize{}$1.11\times10^{-1}\mathcal{C}$} & {\scriptsize{}-} & {\scriptsize{}$2.11\times10^{-2}\mathcal{C}$}\tabularnewline
{\scriptsize{}$L_{\k}$} & {\scriptsize{}$L_{\k}^{*}/L^{*}$} & {\scriptsize{}\makecell[l]{Ratio of region thickness \\ to cell thickness} } & {\scriptsize{}0.493} & {\scriptsize{}$6.94\times10^{-2}$} & {\scriptsize{}0.438}\tabularnewline
{\scriptsize{}$\sigma_{\k}$} & {\scriptsize{}$(R^{*}T^{*}/F^{*})/(i_{\mathrm{typ}}^{*}L^{*}/\sigma_{\k}^{*})$} & {\scriptsize{}\makecell[l]{Ratio of thermal voltage to \\ the typical Ohmic drop in the solid} } & {\scriptsize{}$6.39\times10^{3}/\mathcal{C}$} & {\scriptsize{}-} & {\scriptsize{}$5.35/\mathcal{C}$}\tabularnewline
{\scriptsize{}$a_{\mathrm{tot},\k}$} & {\scriptsize{}$\bar{R}_{\k,a}^{*}a_{\mathrm{tot},\k}^{*}$} & {\scriptsize{}\makecell[l]{Product of mean radius and \\ active surface area per volume. \\
(By choice of mean radius, \\ this also corresponds to $3 \epsilon_{\s,\k}$.)} } & {\scriptsize{}2.25} & {\scriptsize{}-} & {\scriptsize{}1.99}\tabularnewline
{\scriptsize{}$\gamma_{\k}$} & {\scriptsize{}$c_{\k,\mathrm{max}}^{*}/c_{\n,\mathrm{max}}^{*}$} & {\scriptsize{}\makecell[l]{Maximum lithium concentration \\ in solid relative to maximum \\ in negative electrode} } & {\scriptsize{}1} & {\scriptsize{}-} & {\scriptsize{}1.90}\tabularnewline
{\scriptsize{}$c_{\k,0}$} & {\scriptsize{}$c_{\k,0}^{*}/c_{\k,\mathrm{max}}^{*}$} & {\scriptsize{}Initial stoichiometry in active material.} & {\scriptsize{}0.901} & {\scriptsize{}-} & {\scriptsize{}0.270}\tabularnewline
{\scriptsize{}$\mathcal{C}_{\e}$} & {\scriptsize{}$\tau_{\e}^{*}/\tau_{\mathrm{d}}^{*}$} & {\scriptsize{}\makecell[l]{Ratio of electrolyte diffusion \\ to discharge timescales}} & \multicolumn{3}{c}{{\scriptsize{}$1.53\times10^{-3}\mathcal{C}$}}\tabularnewline
{\scriptsize{}$\gamma_{\e}$} & {\scriptsize{}$c_{\e,0}^{*}/c_{\n,\mathrm{max}}^{*}$} & {\scriptsize{}\makecell[l]{Typical lithium concentration \\ in electrolyte relative to \\ maximum in negative electrode} } & \multicolumn{3}{c}{{\scriptsize{}$3.02\times10^{-2}$}}\tabularnewline
{\scriptsize{}$\hat{\kappa}_{\e}$} & {\scriptsize{}$(R^{*}T^{*}/F^{*})/(i_{\mathrm{typ}}^{*}L^{*}/\kappa_{\e,\mathrm{typ}}^{*})$} & {\scriptsize{}\makecell[l]{Ratio of thermal voltage to \\ the typical Ohmic drop \\ in the electrolyte} } & \multicolumn{3}{c}{{\scriptsize{}$2.82\times10^{1}/\mathcal{C}$}}\tabularnewline
\hline 
\end{tabular}
\par\end{centering}
\caption{Dimensionless parameters and their definitions in terms of dimensional
(asterisks) parameters. Dependence on the C-rate, denoted $\mathcal{C}$,
shown explicitly.}
\label{tab:Dimensionless-parameters}
\end{table*}

\begin{table*}
\begin{centering}
\begin{tabular}{lrlllc}
\hline 
 & \multirow{2}{*}{{\scriptsize{}Variable}} & \multirow{2}{*}{{\scriptsize{}Scaling}} & \multirow{2}{*}{{\scriptsize{}Description}} & {\scriptsize{}Depends on} & \multirow{2}{*}{{\scriptsize{}Regions $\k$}}\tabularnewline
 &  &  &  & {\scriptsize{}coordinates} & \tabularnewline
\hline 
\hline 
\multirow{4}{*}{{\scriptsize{}Coordinates}} & {\scriptsize{}$t$} & {\scriptsize{}$t^{*}/\tau_{\mathrm{d}}^{*}$} & {\scriptsize{}Time} & {\scriptsize{}-} & {\scriptsize{}-}\tabularnewline
 & {\scriptsize{}$x$} & {\scriptsize{}$x^{*}/L^{*}$} & {\scriptsize{}\makecell[l]{Distance from negative \\ current collector}} & {\scriptsize{}-} & {\scriptsize{}-}\tabularnewline
 & {\scriptsize{}$R_{\k}$} & {\scriptsize{}$R_{\k}^{*}/\bar{R}_{\k,a}^{*}$} & {\scriptsize{}Active particle radius} & {\scriptsize{}-} & {\scriptsize{}$\n,\p$}\tabularnewline
 & {\scriptsize{}$r_{\k}$} & {\scriptsize{}$r_{\k}^{*}/R_{\k}^{*}$} & {\scriptsize{}Radial coordinate in active particle} & {\scriptsize{}-} & {\scriptsize{}$\n,\p$}\tabularnewline
\hline 
\multirow{4}{*}{{\scriptsize{}\makecell[l]{Fundamental \\ variables}}} & {\scriptsize{}$\phi_{\s,\k}$} & {\scriptsize{}$(\phi_{\s,\k}^{*}-\phi_{\s,\k}^{*}|_{t^{*}=0})F^{*}/(R_{g}^{*}T^{*})$} & {\scriptsize{}Electric potential in the solid} & {\scriptsize{}$t,x$} & {\scriptsize{}$\n,\p$}\tabularnewline
 & {\scriptsize{}$\phi_{\e,\k}$} & {\scriptsize{}$(\phi_{\e,\k}^{*}-\phi_{\e,\k}^{*}|_{t^{*}=0})F^{*}/(R_{g}^{*}T^{*})$} & {\scriptsize{}Electric potential in the electrolyte} & {\scriptsize{}$t,x$} & {\scriptsize{}$\n,\sep,\p$}\tabularnewline
 & {\scriptsize{}$c_{\s,\k}$} & {\scriptsize{}$c_{\s,\k}^{*}/c_{\k,\mathrm{max}}^{*}$} & {\scriptsize{}\makecell[l]{Lithium concentration \\ in the active material}} & {\scriptsize{}$t,x,R_{\k},r_{\k}$} & {\scriptsize{}$\n,\p$}\tabularnewline
 & {\scriptsize{}$c_{\e,\k}$} & {\scriptsize{}$c_{\e,\k}^{*}/c_{\e,0}^{*}$} & {\scriptsize{}\makecell[l]{Lithium-ion concentration \\ in the electrolyte}} & {\scriptsize{}$t,x$} & {\scriptsize{}$\n,\sep,\p$}\tabularnewline
\hline 
\multirow{11}{*}{{\scriptsize{}\makecell[l]{derived \\ (known in terms \\ of fundamental)}}} & {\scriptsize{}$i_{\s,\k}$} & {\scriptsize{}$i_{\s,\k}^{*}/i_{\mathrm{typ}}^{*}$} & {\scriptsize{}Current density in the solid} & {\scriptsize{}$t,x$} & {\scriptsize{}$\n,\p$}\tabularnewline
 & {\scriptsize{}$i_{\e,\k}$} & {\scriptsize{}$i_{\e,\k}^{*}/i_{\mathrm{typ}}^{*}$} & {\scriptsize{}Current density in the electrolyte} & {\scriptsize{}$t,x$} & {\scriptsize{}$\n,\sep,\p$}\tabularnewline
 & {\scriptsize{}$N_{\s,\k}$} & {\scriptsize{}$R_{\k}^{*}N_{\s,\k}^{*}/(D_{\s,\k}^{*}c_{\k,\mathrm{max}}^{*})$} & {\scriptsize{}Lithium flux in the active material} & {\scriptsize{}$t,x$} & {\scriptsize{}$\n,\p$}\tabularnewline
 & {\scriptsize{}$N_{\e,\k}$} & {\scriptsize{}$L^{*}N_{\e,\k}^{*}/(D_{\e,\mathrm{typ}}^{*}c_{\e,0}^{*})$} & {\scriptsize{}Lithium-ion flux in the electrolyte} & {\scriptsize{}$t,x$} & {\scriptsize{}$\n,\sep,\p$}\tabularnewline
 & {\scriptsize{}$J_{\k}$} & {\scriptsize{}$J_{\k}^{*}L^{*}/i_{\mathrm{typ}}^{*}$} & {\scriptsize{}Total interfacial current density} & {\scriptsize{}$t,x$} & {\scriptsize{}$\n,\p$}\tabularnewline
 & {\scriptsize{}$j_{\k}$} & {\scriptsize{}$j_{\k}^{*}a_{\mathrm{tot},\k}^{*}L^{*}/i_{\mathrm{typ}}^{*}$} & {\scriptsize{}Interfacial current density} & {\scriptsize{}$t,x,R_{\k}$} & {\scriptsize{}$\n,\p$}\tabularnewline
 & {\scriptsize{}$j_{0,\k}$} & {\scriptsize{}$j_{0,\k}^{*}a_{\mathrm{tot},\k}^{*}L^{*}/i_{\mathrm{typ}}^{*}$} & {\scriptsize{}Interfacial exchange current density} & {\scriptsize{}$t,x,R_{\k}$} & {\scriptsize{}$\n,\p$}\tabularnewline
 & {\scriptsize{}$\eta_{\k}$} & {\scriptsize{}$F^{*}\eta_{\k}^{*}/(R_{g}^{*}T^{*})$} & {\scriptsize{}Reaction overpotential} & {\scriptsize{}$t,x,R_{\k}$} & {\scriptsize{}$\n,\p$}\tabularnewline
 & {\scriptsize{}$U_{\k}$} & {\scriptsize{}$(U_{\k}^{*}-U_{\k,0}^{*})F^{*}/(R_{g}^{*}T^{*})$} & {\scriptsize{}Open circuit potential} & {\scriptsize{}$t,x,R_{\k}$} & {\scriptsize{}$\n,\p$}\tabularnewline
 & {\scriptsize{}$f_{\k,a}$} & {\scriptsize{}$\bar{R}_{\k,a}^{*}f_{\k,a}^{*}$} & {\scriptsize{}\makecell[l]{Area-weighted particle-size \\ distribution of active material \\ (mean 1)}} & {\scriptsize{}$R_{\k}$} & {\scriptsize{}$\n,\p$}\tabularnewline
 & {\scriptsize{}$D_{\e}$} & {\scriptsize{}$D_{\e}^{*}/D_{\e,\mathrm{typ}}^{*}$} & {\scriptsize{}\makecell[l]{Lithium-ion diffusivity \\ in the electrolyte}} & {\scriptsize{}$t,x$} & {\scriptsize{}$\n,\sep,\p$}\tabularnewline
\hline 
\end{tabular}
\par\end{centering}
\caption{Dimensionless variables, the relations to their corresponding dimensional
(asterisk) variable, and their regions of definition. The values correspond
to the parameter set from \citep{Chen2020}, given in Table \ref{tab:Dimensional-parameters}. }
\label{tab:Dimensionless-variables}
\end{table*}

\subsubsection*{Charge conservation}

\begin{align}
\frac{\partial i_{\e,\k}}{\partial x} & =\begin{cases}
J_{\k}, & \k=\n,\p,\\
0, & \k=\sep,
\end{cases} & \k\in\{\n,\sep,\p\},\label{eq:charge_conservation}\\
i_{\s,\k} & =i_{\mathrm{app}}(t)-i_{\e,\k}, & \k\in\{\n,\p\}
\end{align}
\begin{equation}
i_{\e,\k}=\epsilon_{\k}^{b}\hat{\kappa}_{\e}\kappa_{\e}(c_{\e,\k})\left[-\frac{\partial\phi_{\e,\k}}{\partial x}+2(1-t^{+})\frac{\partial}{\partial x}\log c_{\e,\k}\right],\qquad\k\in\{\n,\sep,\p\},
\end{equation}
\begin{equation}
i_{\s,\k}=-\sigma_{\k}\frac{\partial\phi_{\s,\k}}{\partial x},\qquad\k\in\{\n,\p\},
\end{equation}
At $x=L_{\n},1-L_{\p}$ there is continuity of $i_{\e,\k},\phi_{\e,\k}$
and $i_{\s,\k}=0$, and at the current collectors,
\begin{align}
i_{\e,\k} & =0,\qquad\text{at }x=0,1\\
i_{\s,\k} & =i_{\mathrm{app}}(t),\qquad\text{at }x=0,1\\
\phi_{\s,\n} & =0\qquad\text{at }x=0,\\
\phi_{\s,\p} & =V(t)\qquad\text{at }x=1.
\end{align}

\subsubsection*{Molar conservation of lithium}

\begin{align}
\mathcal{C}_{\e}\gamma_{\e}\epsilon_{\k}\frac{\partial c_{\e,\k}}{\partial t} & =-\gamma_{\e}\frac{\partial N_{\e,\k}}{\partial x}+\mathcal{C}_{\e}\frac{\partial i_{\e,\k}}{\partial x},\quad\k\in\{\n,\sep,\p\},\label{eq:dimensional_lithium_cons_electrolyte-1}\\
N_{\e,\k} & =-\epsilon_{\k}^{b}D_{\e}(c_{\e,\k})\frac{\partial c_{\e,\k}}{\partial x}+\frac{\mathcal{C}_{\e}t^{+}}{\gamma_{\e}}i_{\e,\k},\quad\k\in\{\n,\sep,\p\},
\end{align}
At $x=L_{\n},1-L_{\p}$ there is continuity of $c_{\e,\k},N_{\e,\k}$,
and at $x=0,1$ there is no-flux, $N_{\e,\k}=0$. 

\begin{align}
\mathcal{C}_{\k}R_{\k}\frac{\partial c_{\s,\k}}{\partial t} & =-\frac{1}{r_{\k}^{2}}\frac{\partial}{\partial r_{\k}}(r_{\k}^{2}N_{\s,\k}),\quad\k\in\{\n,\p\},\\
N_{\s,\k} & =-\frac{\partial c_{\s,\k}}{\partial r_{\k}},\quad\k\in\{\n,\p\},
\end{align}
\begin{align}
N_{\s,\k} & =0,\quad\text{at }r_{\k}=0, & \frac{a_{\k}\gamma_{\k}}{\mathcal{C}_{\k}R_{\k}}N_{\s,\k} & =j_{\k},\quad\text{at }r_{\k}=1,
\end{align}
and
\begin{equation}
J_{\k}=\int_{\Omega_{\k}^{'}}f_{\k,a}(R_{\k})j_{\k}\,\mathrm{d}R_{\k},\qquad\k\in\{\n,\p\}.\label{eq:dimensional_J_k-1}
\end{equation}

\subsubsection*{Electrochemical reactions}

At $r_{\k}=1$,
\begin{align}
j_{\k} & =j_{\k,0}\sinh\left(\frac{\eta_{\k}}{2}\right),\qquad\k\in\{\n,\p\},\\
j_{0,\k} & =\frac{\gamma_{\k}}{\mathcal{C}_{\mathrm{r},\k}}(c_{\s,\k})^{1/2}(1-c_{\s,\k})^{1/2}(c_{\e,\k})^{1/2},\qquad\k\in\{\n,\p\},\\
\eta_{\k} & =\phi_{\s,\k}-\phi_{\e,\k}-U_{\k}(c_{\s,\k}),\qquad\k\in\{\n,\p\},
\end{align}

\subsubsection*{Initial conditions}

At $t=0$,
\begin{align}
c_{\s,\k} & =c_{\k,0},\qquad\k\in\{\n,\p\},\\
c_{\e,\k} & =1,\qquad\k\in\{\n,\sep,\p\},
\end{align}
and all other variables initially equal to zero.

\bibliographystyle{unsrt}
\bibliography{Refdatabase8}

\end{document}